\documentclass[aps,prb,twocolumn,superscriptaddress,showpacs,10pt]{revtex4-1}
\usepackage{natbib}
\usepackage{mathbbol,amsmath,amsfonts,amssymb,bm,color,dsfont,bbold}
\usepackage{graphicx,psfrag,array,braket}

\providecommand{\ignore}[1]{}
\providecommand{\aucmnt}[1]{#1}
\def\be{\begin{equation}}
\def\ee{\end{equation}}
\renewcommand{\aucmnt}[1]{}

\newcommand{\Comment}[1]{}

\newcommand{\Eq}[1]{Eq.~(\ref{#1})}

\begin{document}

\title{The Jain-2/5 parent Hamiltonian: structure of zero modes, dominance patterns, and zero mode generators}
\author{Li Chen}
\affiliation{Department of Physics, Washington University, St.
Louis, MO 63130, USA}
\affiliation{National High Magnetic Field Laboratory and Department of Physics, Florida State University, Tallahassee, FL 32306, USA}
\author{Sumanta Bandyopadhyay}
\author{Alexander Seidel}
\affiliation{Department of Physics, Washington University, St.
Louis, MO 63130, USA}

\date{\today}
\begin{abstract}
We analyze general zero mode properties of the parent Hamiltonian of the unprojected Jain-2/5 state. We characterize the zero mode condition associated to this Hamiltonian via projection onto a four-dimensional two-particle subspace for given pair angular momentum, for the disk and similarly for the spherical geometry. Earlier numerical claims in the literature about ground state uniqueness on the sphere are substantiated on analytic grounds, and related results are derived. Preference is given to second quantized methods, where zero mode properties are derived not from given analytic wave functions, but from a ``lattice'' Hamiltonian and associated zero mode conditions. This method reveals new insights into the guiding-center structure of the unprojected Jain-2/5 state, in particular a system of dominance patterns following a ``generalized Pauli principle'', which  establishes a complete one-to-one correspondence with the edge mode counting. We also identify one-body operators that function as generators of zero modes.
\end{abstract}
\maketitle

\section{Introduction}

The theoretical exploration of topological phases in the fractional quantum Hall (FQH) regime owes its success to our ability to associate simple data to valid points in the phase diagram. At the level of the low-energy effective theory, these data may be thought of describing a topological quantum field theory, or a related rational conformal field theory. Remarkably, the same data lend themselves to the construction of microscopic many-body wave functions.\cite{MR} In this context, such data have also been thought of as ``dancing patterns''.\cite{wen_toprev}
To the extent that these patterns translate into simple analytic properties of wave functions, often the construction of a local parent Hamiltonian is also possible. This situation may be thought of as nearly ideal: The existence of simple data that both lead to an effective field theory as well as a solvable microscopic Hamiltonian. This last step, however, the construction of a Hamiltonian, has not always been successful thus far.
The experimentally most important sequence of quantum Hall states appears to be described by weakly interacting composite fermions as originally discussed by Jain.\cite{Jain}
This includes states described by Laughlin's seminal wave functions,\cite{Laughlin}
for which parent Hamiltonians have been successfully constructed early on.\cite{haldane_hierarchy, TK}
However, for the majority of Jain states, there seem no successful attempts at construction of a parent Hamiltonian thus far. To the best of our knowledge, this is in particular true for all lowest Landau-level projected versions of Jain states, aside from Laughlin states.

A special niche seems to be occupied by the {\em un-}projected Jain-2/5 state. Generally speaking, the fact that the polynomial part of the wave function is no longer holomorphic, but depends on both holomorphic and anti-holomorphic complex coordinates when higher Landau levels are involved, tends to make it more difficult to identify analytic clustering principles that allow for the construction of a Hamiltonian. However, for the unprojected Jain-2/5 state a parent Hamiltonian has been identified.\cite{ReMac}
Well-studied parent Hamiltonians in the fractional quantum Hall regime tend to achieve more than just stabilizing a ground state: The ground state is the unique zero energy state (zero mode) at a given filling factor, but is degenerate with other zero modes when the filing factor is reduced by introducing more flux quanta or reducing particle number. The number of these additional zero modes at fixed angular momentum relative to the ground state is generally in one-to-one correspondence with edge mode counting in the conformal field theory describing the edge.\cite{Read_edgetheory} Here, the angular momentum of the microscopic zero mode, relative to the ground state plays the role of energy in the effective edge theory, as may be justified by adding a confining potential proportional to angular momentum.
We will say that a Hamiltonian that conforms to the above paradigm satisfies the ``zero mode paradigm''.
For many quantum Hall parent Hamiltonians involving projection onto the lowest Landau level, pertinent zero mode counting exercises have a long tradition in the field.\cite{RRcount, ardonne,ArdonneJPhysA35:447, SY11}
However, the parent Hamiltonian of the Jain-2/5 state involves projection onto two (artificially quenched) Landau levels. Here, the situation seems to have been less studied. One of the results of this paper will be the rigorous characterization of all zero modes of this Hamiltonian and their one-to-one correspondence with degrees of freedom  the edge theory, including certain ``zero-momentum modes'' of the latter that involve changes in particle number or transfer of particles between different edge branches.

Moreover, recent years have shown that the data specifying a topological phase in the fractional quantum Hall regime largely survives in certain skeletal forms of special wave functions associated with the thin torus limit \cite{seidel05,SL,SY08, BK1,BK2,BK3}
or with ``dominant partitions''.\cite{BH1, BH2,BH3,ThomalePRB84:45127}
These in particular contain information about quasiparticle statistics (see \onlinecite{Flavin} for a review).
Furthermore, in Ref. \onlinecite{ortiz}
a mechanism was identified that explains the appearance of such dominant partitions, or dominance patterns, in any quantum Hall wave function for which a parent Hamiltonian with the properties described above can be given.
It is worth noting that this is different from relating such dominance patterns to analytic clustering properties of first quantized wave functions, which  was done by the original work.\cite{BH1,BH2,BH3, ThomalePRB84:45127}
While the latter approach does not utilize a Hamiltonian principle of the kind described above (which may not always be available), the approach pursued here and in earlier works by some of us does not require analytic clustering conditions (which may not always be present\cite{Nakamura1, Nakamura2}).
Moreover, we will argue below that the Hamiltonian approach may give some insights into why certain types of wave functions cannot be stabilized by a Hamiltonian satisfying the zero mode paradigm. In particular, we will argue this to be the case for projected Jain states.

Lastly, it appears that neither approach to dominance patterns has so far been applied to a situation where the many-body state was not described by holomorphic wave functions (modulo non-holomorphic, e.g., Gaussian factors common to all states).
In the majority of cases, dominance patterns for special FQH wave functions have been discussed for single component states in the lowest Landau level. In some cases, additional degrees of freedom such as spin were present.\cite{SY1, SY2} Also, Landau level projected Jain states have been discussed from the point of view of dominance patterns.\cite{RBH} The idea of this paper is to present a case study for both the zero mode paradigm as well as a description in terms of dominance patterns, and the interplay between these concepts, for a state that is not projected onto the lowest Landau level, and whose wave function is consequently not holomorphic. For present purposes, this will be the unprojected Jain-2/5 state.

The remainder of the paper is organized as follows. In Sec. \ref{2ndq}, we will present the second quantized form of the parent Hamiltonian of this state on the disk, which represents the natural framework for our approach. In Sec. \ref{dominance}, we will use this second quantized form  to establish a description of zero modes in terms of dominance patterns.
In Sec. \ref{countedge}, we use these results to establish the one-to-one correspondence between zero modes, dominance patterns, and modes of the edge theory. In Sec. \ref{zmgn},
we present second quantized single particle operators that serve as generators for zero modes. In Sec. \ref{sphere} we extend our main results to the spherical geometry. In Sec. \ref{discussion}, we will discuss our results. We conclude in Sec. \ref{conclusion}.

\section{Second quantization in disk geometry\label{2ndq}}

In this paper, we will be concerned with the two-body Trugman-Kivelson interaction\cite{TK} \be\label{H}
  H =P_n\, \nabla _1^2\delta \left( {{x_1} - {x_2}} \right)\delta \left( {{y_1} - {y_2}} \right)P_n\,,
\ee
projected onto the first $n$ Landau levels via an orthogonal projection operator $P_n$, focusing on the case where $n=2$.
For $n=1$, it is well known that this interaction agrees, up to a factor, with the $V_1$ Haldane
pseudopotential.\cite{haldane_hierarchy}
The case $n=2$ was identified by Rezayi and MacDonald \cite{ReMac} as a
parent Hamiltonian for the Jain-2/5 state, where at the same time, the kinetic energy is quenched not only within individual Landau levels, but the splitting between the lowest and first excited Landau level is set to zero. Here we will mainly be concerned with the properties of this ($n=2$) Hamiltonian.
Results for the case $n=3$ have appeared  recently.\cite{wu2016new}  The extension of the methods developed below to $n>2$ is left to a forthcoming paper.


As a starting point, we establish a second quantized form of the Hamiltonian in various geometries, beginning with the disk geometry.
For positive, angular momentum conserving two-particle operators, the second quantized many-body Hamiltonian is generally\cite{ortiz} of the form
\be \label{H2nd}
 H= \sum_{k=1}^M \sum_R {T^{(k)}_R}^\dagger {T^{(k)}_R}\,,
\ee
where ${T^{(k)}_R}=\sum_x f^k_{i,j}(R,x) c_{i,R-x} c_{j,R+x}$ destroys a pair of particles with well defined angular momentum $2R$, $c_{i,m}$ is an electron destruction operator for a state in the $i$th Landau level (LL) with angular momentum $m$, and $f^k_{i,j}(R,x)$ is a form factor defining the operator $T^{(k)}_R$. In \Eq{H2nd}, The sum over $R$ is over integer and half-odd integer values,
and $x$ in the definition of $T^{(k)}_R$ is {\em either} over integer {\em or} half-odd integer, depending on $R$ (i.e., $2x\equiv 2R \mod 2$). In the most general case, the number $M$ of families of $T$-operators can be infinite.

We now work out the connection between Eqs. \eqref{H} and \eqref{H2nd} specializing to $n=2$ Landau levels (carrying Landau level indices $0$ and $1$, respectively). To this end, we recall  the wave functions for a single particle in the disk with angular momentum $L_z=m$ in the lowest and first excited LLs under symmetric gauge,
\be
  \eta_{0,m}(z) = \frac{{z^m}{e^{ - {|z|^2}/4l_B^2}}}{{\sqrt {2\pi {2^m}l_B^{2m + 2}m!} }}
\ee
and
\be
  \eta_{1,m}(z) = \frac{\left( {\bar{z} {z^{m+1}} - 2l_B^2(m+1)z^m} \right){e^{ - {|z|^2}/4l_B^2}}}{{\sqrt {2\pi {2^{m+2}}l_B^{2m+6}(m+1)!} }},
\ee
respectively, where $z=x+i y$ is the complex coordinate on the disk, and $l_B$ is magnetic length $\sqrt{\hbar/eB}$.
As an immediate consequence, we have the following analytic structure for general {\em two-particle}
wave functions projected onto the first two LLs,
\be\label{psi}
\begin{split}
\psi(z_1,z_2)=&\Big(C_{00}(z_1,z_2)+\bar{z}_1C_{10}(z_1,z_2)+\bar{z}_2C_{01}(z_1,z_2)\\
&+\bar{z}_1\bar{z}_2C_{11}(z_1,z_2)\Big)e^{-\frac{|z_1|^2}{4l_B^2}-\frac{|z_2|^2}{4l_B^2}},
\end{split}
\ee
where $C_{00}(z_1,z_2), C_{10}(z_1,z_2), C_{01}(z_1,z_2)$ and $ C_{11}(z_1,z_2)$ are holomorphic functions of $z_1$ and $z_2$.
For two-particle states, it is  generally advantageous to phrase expressions in terms
of a center-of-mass coordinate $z_c=(z_1+z_2)/2$ and a relative coordinate $z_r=z_1-z_2$,
and their complex conjugates $\bar z_c$, $\bar z_r$.
Furthermore, in this paper we will be exclusively considering fermions.
 Then, \Eq{psi} can be recast as
\be\label{psai}
\begin{split}
\psi(z_c,z_r)=&\Big(d_{00}(z_c,z_r)+\bar{z}_c d_{10}(z_c,z_r)+\bar{z}_r d_{01}(z_c,z_r)\\
&+(\bar z_c^2 - \bar z_r^2/4) d_{11}(z_c,z_r)\Big)e^{-\frac{|z_c|^2}{2l_B^2}-\frac{|z_r|^2}{8l_B^2}},
\end{split}
\ee
where $d_{00}(z_c,z_r), d_{10}(z_c,z_r), d_{01}(z_c,z_r)$ and $ d_{11}(z_c,z_r)$ are holomorphic functions of $z_r$ and $z_c$ with well-defined parity in $z_r$. Specifically,  antisymmetry dictates that $d_{00}(z_c,z_r), d_{10}(z_c,z_r), d_{11}(z_c,z_r)$ are odd in $z_r$ whereas $d_{01}(z_c,z_r)$ is even in $z_r$.
It will be beneficial to work with an orthogonal basis of two-particle states that preserves
as far as possible a factorization into center-of-mass and relative parts. Note that unlike the lowest LL,
higher Landau levels are {\em not} invariant subspaces of the relative or center-of mass angular
 momentum operators individually, hence unlike in the lowest LL, there are no good quantum numbers associated with these observables. This is related to the presence of the last term in \Eq{psai}.
 We thus write:
 \be\begin{split}\label{ortho}
 \psi(z_c,z_r)=\sum_{R,\ell} \left\{
 a_{R,\ell} \,\eta_{0,\ell}^r(z_r)\eta_{0,2R-\ell}^c(z_c)+\right . \\
 b_{R,\ell} \,\eta_{0,\ell}^r(z_r)\eta_{1,2R-\ell}^c(z_c)+\\
 c_{R,\ell} \,\eta_{1,\ell}^r(z_r)\eta_{0,2R-\ell}^c(z_c)+\\
\left. d_{R,\ell} \,\left(\eta_{0,\ell}^r(z_r)\eta_{2,2R-\ell}^c(z_c)-\eta_{2,\ell-2}^r(z_r)\eta_{0,2R+2-\ell}^c(z_c)\right)/\sqrt{2}
  \right\} ,
 \end{split}
 \ee
where functions $\eta^r_{k,m}(z_r)$ and $\eta^c_{k,m}(z_c)$ are obtained from $\eta_{k,m}(z)$ via
$l_B\rightarrow\sqrt{2} l_B$ and $l_B\rightarrow l_B/\sqrt{2}$, respectively, $\ell$ is restricted to odd integers, the $k=0,1$
Landau level wave functions were given above, and those for $k=2$ are also needed:
\be
\begin{split}
 &\eta_{2,m}(z) = {e^{ - {|z|^2}/4l_B^2}}\\&\times\frac{z^m({\bar{z}^2} {z^{2}}- 4l_B^2(m+2) \bar{z} {z}+4l_B^4(m+2)(m+1) )}{{\sqrt {2\pi {2^{m+5}}l_B^{2m+10}(m+2)!} }} .  \end{split}
\ee
It is easy to see that \Eq{ortho} reproduces the analytic structure of \Eq{psai}.
Moreover, for sufficiently rapidly decaying $\psi(z_c,z_r)$, which we will always assume,
any such $\psi(z_c,z_r)$ can be expanded in the form \Eq{ortho}, which
follows from completeness
properties of the $\eta$-functions.

One may see that the Hamiltonian \Eq{H} is positive (semi-definite) for general $  n$,
which will be made explicit for $n=2$ below. Therefore, as in the more familiar case $n=1$,
any zero modes are exact ground states. One may further
see easily that the familiar analyticity requirements for zero modes for $n=1$ generalize
as follows.  For the two-particle state \eqref{ortho} {\em not} to be annihilated by $H$
(i.e., to have any non-zero matrix elements within the image of $P_n$), its polynomial
expansion (not including the Gaussian term) must have terms that are at most linear
in $z_r$, $\bar z_r$. 
With this in mind, working at fixed angular momentum $L_z=2R$ at the moment, we see
that all non-zero eigenstates of $H$ must be contained in the six-dimensional subspace spanned
by the following states,

\begin{subequations}\label{S16}
 \begin{eqnarray}\eta^r _{1, - 1}({z_r} )\eta^c _{0,2R + 1}\left( {{z_c}} \right),\end{eqnarray}
\begin{eqnarray}\eta^r _{0,1}\left( {z_r} \right)\eta^c _{0,2R - 1}\left( {{z_c}} \right),\end{eqnarray}
\begin{eqnarray}
\eta^r _{0,1}\left( {z_r} \right)\eta^c _{1,2R - 1}\left( {{z_c}} \right),\end{eqnarray}
\begin{eqnarray}
\frac{\left( \eta^r _{0,1}\left( {z_r} \right)\eta^c _{2,2R - 1}\left( {{z_c}} \right)- \eta^r _{2, - 1}\left( {z_r} \right)\eta^c _{0,2R + 1}(z_c) \right)}{\sqrt 2},\end{eqnarray}
\begin{eqnarray}
\eta^r _{1,1}\left( {z_r} \right)\eta^c _{0,2R - 1}\left( {{z_c}} \right),\end{eqnarray}
\begin{eqnarray}
\frac{\left( {\eta^r _{0,3}\left( {z_r} \right)\eta^c _{2,2R - 3}\left( {{z_c}} \right)- \eta^r _{2,1}\left( {z_r} \right)\eta^c _{0,2R - 1}\left( {{z_c}} \right)} \right)}{\sqrt 2}\,,\end{eqnarray}
\end{subequations}
while its orthogonal complement (for given $R$) is spanned by states already annihilated by $H$.
It follows from this that the Hamiltonian may be written in the form
\be\label{HQ1}
   H=\sum_R\sum_{i,j=1}^6 m_{i,j} {Q^{(i)}_R}^\dagger Q^{(j)}_R\,
\ee
where the operators ${Q^{(i)}_R}^\dagger$, $i=1\dotsc 6$, create the states in \Eq{S16}.
Specifically, in second quantized form, these operators read:
\begin{subequations}\label{Q16}
\begin{eqnarray}
Q_R^{(1)} = \frac{1}{2^{R + 1/2}}\sum\limits_{x =  - R}^{R + 1} {\sqrt {\left( \begin{gathered}
  2R + 1 \\
  R + x \\
\end{gathered}  \right)} } {c_{1, R - x}}{c_{0, R + x}},\end{eqnarray}
\begin{eqnarray}
Q_R^{(2)} =  - \frac{1}{2^R}\sum\limits_{x =  - R}^R {x\sqrt {\frac{1}{R}\left( \begin{gathered}
  2R \\
  R + x \\
\end{gathered}  \right)} } {c_{0, R - x}}{c_{0, R + x}},\end{eqnarray}\,
\be
\begin{split}
Q_R^{(3)} = \frac{1}{2^{R + 1/2}}\sum\limits_{x =  - R}^{R + 1} &\left( {1 - 2x} \right)\sqrt {\frac{1}{2R + 1}\binom{2R + 1}{R + x }} \\ &\times c_{1, R - x}c_{0, R + x},\end{split}
\ee
\be
\begin{split}
Q_R^{(4)} =   - \frac{1}{2^{R + 1/2}}\sum\limits_{x =  - R - 1}^{R + 1} &{x\sqrt {\frac{1}{{2R + 2}}\left( \begin{gathered}
  2R + 2 \\
  R + 1 + x \\
\end{gathered}  \right)} } \\ &\times{c_{1, R - x}}{c_{1, R + x}},\end{split}\ee
\be
\begin{split}
Q_R^{(5)} =  \frac{1}{2^R}\sum\limits_{x =  - R}^{R + 1} &{\left( {2{x^2} - 2x - R} \right)\sqrt {\frac{1}{{2R\left( {2R + 1} \right)}}\left( \begin{gathered}
  2R + 1 \\
  R + x \\
\end{gathered}  \right)} }\\ &\times {c_{1, R  - x}}{c_{0, R + x}},\end{split}\ee
\be
\begin{split}
Q_R^{(6)}=-\frac{1}{2^R\sqrt 3}&\sum\limits_{x =-R-1}^{R+1}(2{x^3} - (3R+2)x)
\\ & \times \sqrt {\frac{1}{2R(2R+1)(2R+2)}\binom{2R+2} {R+1+x}}\\ & \times c_{1,R-x}c_{1,R+x}.
\end{split}\ee
\end{subequations}
As before, $x$ is summed over (half)integers when $R$ is  (half)integer. Possible
values for $R \pm x$ are non-negative for Landau level index  $i=0$, and are greater than or equal to $-1$ for $i=1$, to accommodate for the $L_z=-1$ angular momentum state in the first excited Landau level.
One may check that these operators satisfy
$\bra 0 Q_R^{(n)}{Q_{R'}^{(m)}}^\dag  \ket 0 = \delta _{n,m}\delta _{R,R'}$, as expected from the orthonormality of first quantized wave functions used in this analysis.
The matrix elements $m_{ij}$ in \Eq{HQ1} turn out to be independent of $R$, and can be read of the following expression:
\be
\begin{split}
  H &=  \frac{1}{{4\pi }}{\sum\limits_R {Q_R^{(1)}} ^\dag }Q_R^{(1)} + \frac{3}{{8\pi }}{\sum\limits_R {Q_R^{(4)}} ^\dag }Q_R^{(4)} \\& + \frac{1}{{4\pi }}{\sum\limits_R ({Q_R^{(1)}} ^\dag }Q_R^{(4)}+\text{h.c.})+\frac{1}{{4\pi }}{\sum\limits_R {Q_R^{(3)}} ^\dag }Q_R^{(3)}
  \\&+ \frac{1}{{4\pi }}{\sum\limits_R {Q_R^{(2)}} ^\dag }Q_R^{(2)} + \frac{1}{{2\pi }}{\sum\limits_R {Q_R^{(5)}} ^\dag }Q_R^{(5)}
  \\& +\frac{3}{{8\pi }}{\sum\limits_R {Q_R^{(6)}} ^\dag }Q_R^{(6)} - \frac{{\sqrt 2 }}{{4\pi }}{\sum\limits_R ({Q_R^{(2)}} ^\dag }Q_R^{(5)}+\text{h.c.})
  \\& -\frac{{\sqrt 6 }}{{8\pi }}{\sum\limits_R ({Q_R^{(2)}} ^\dag }Q_R^{(6)}+\text{h.c.})+ \frac{{\sqrt 3 }}{{4\pi }}{\sum\limits_R ({Q_R^{(5)}} ^\dag }Q_R^{(6)}+\text{h.c.}) .
\end{split}
\ee

It further turns out that only four of the six eigenvalues of the $m$-matrix are non-zero, having values
 $\frac{5\pm\sqrt{17}}{16\pi}$, $\frac{1}{4\pi}$, and $\frac{9}{8\pi}$, respectively. Eigenstates corresponding to these non-zero eigenvalues are: $\frac{\sqrt{2}}{2\sqrt {17\mp\sqrt{17}}}( (-1\pm\sqrt{17}){Q_R^{(1)}}^\dag+4{Q_R^{(4)}}^\dag )\ket 0$, ${Q_R^{(3)}}^\dag \ket 0 $ and $( { - \sqrt 2 {Q_R^{(2)}} ^\dag  + 2{Q_R^{(5)}} ^\dag  +\sqrt 3   {Q_R^{(6)}} ^\dag } )\ket 0/3$. If we denote the latter by ${  T_R^{(1)\dag}}  \ket 0 $, ${  T_R^{(4)\dag}}  \ket 0 $, ${  T_R^{(3)\dag}}  \ket 0 $ and ${  T_R^{(2)\dag}}  \ket 0 $, then the Hamiltonian can be written in diagonal form: \be\label{H2}
\begin{split}
  H &= \frac{5+\sqrt{17}}{16\pi}\sum\limits_R {  T_R^{(1)\dag}} {  T_R^{(1)}}+ \frac{5-\sqrt{17}}{16\pi}\sum\limits_R   T_R^{(4)\dag}   T_R^{(4)}\\&+ \frac{1}{4\pi}\sum\limits_R  T_R^{(3)\dag}   T_R^{(3)} + \frac{9}{8\pi}\sum\limits_R T_R^{(2)\dag}   T_R^{(2)}.
\end{split}
\ee
After rescaling of the $T$-operators, this is of the form \eqref{H2nd} with $M=4$.
The Hamiltonian \eqref{H2} is manifestly the sum of positive (which we will always take to mean semi-definite) terms.
A direct consequence of this is that any zero  mode
of the Hamiltonian \eqref{H2} must be a simultaneous
zero energy eigenstate of each positive term $T_R^{(k)\dag}   T_R^{(k)}$, and, to this end, must be annihilated by each individual operator $T_R^{(k)}$.
Any zero mode $\ket {\psi_0}$ thus obeys the zero mode condition
\be\label{zmc}
   T_R^{(i)}\ket {\psi_0} =0
\ee for $i=1,2,3,4$ and for any integer or half integer $R$. Equivalently,  zero modes are annihilated by $Q_R^{(1)}$, $Q_R^{(4)}$, $Q_R^{(3)}$ and $   T_R^{(2)}$, leading to a slightly more convenient reformulation of the zero mode condition:
\begin{subequations}\label{zero}
\begin{eqnarray}\label{q1}
Q_R^{(1)}\ket {\psi_0}=0,
\end{eqnarray}
\begin{eqnarray}\label{q3}
Q_R^{(3)}\ket {\psi_0}=0,
\end{eqnarray}
\begin{eqnarray}\label{q4}
Q_R^{(4)}\ket {\psi_0}=0,
\end{eqnarray}
\begin{eqnarray}\label{q256}
T_R^{(2)}\ket {\psi_0}=0.
\end{eqnarray}
\end{subequations}
This generalizes the familiar statement for $n=1$ Landau level,
where the $V_1$ Haldane pseudopotential is a two-body projection operator onto states of relative angular momentum $1$.
Presently, for $n=2$, and for given pair angular momentum $2R$, the spectral decomposition of the Trugman-Kivelson interaction involves four two-particle projection operators, each associated to a one dimensional eigenspace spanned by
$T_R^{(i)\dag}  \ket 0 $, $i=1...4$.
Note that it is no longer possible to ascribe definite relative angular momentum quantum numbers to these states.
Note also that the four coefficients in Eq. \eqref{H2} may be replaced with any positive numbers without affecting the zero mode structure of the theory.

\vspace*{+5mm}
\section{Derivation of general properties of dominance patterns in disk geometry\label{dominance}}
With the second quantized form of the parent Hamiltonian, we are now in a position to analyze properties of what we will call general dominance patterns of zero modes of this Hamiltonian. To this end, we will utilize a recently developed method\cite{ortiz} to extract dominance patterns of zero modes directly from the parent Hamiltonian, without any need for studying presupposed wave functions. This has the advantage that since rules for root patterns are arrived at directly as properties of the Hamiltonian, these rules immediately provide rigorous constraints on the zero mode counting for the respective Hamiltonian. In particular, upper bounds for the number of zero modes are immediately available (which we will subsequently show to be saturated), and in particular claims about the unprojected Jain state as the unique densest zero mode of its parent Hamiltonian are immediately established (and in some geometries, refined).
Such claims have appeared earlier in the literature,\cite{ReMac, Wen1991} but, by our reading, have so far been based on numerics, and were thus limited to finite particle number. The present treatment will be free of such limitations.

We begin by clarifying what we mean by a dominance pattern. The notion of a dominance pattern has mainly appeared in the literature in the context of single component states, where dominance patterns are essentially simple product states associated to more complicated quantum Hall trial wave functions. The present situation involves Landau level mixing and is more akin to that in multi-component states, which is more complicated and was described in Refs. \onlinecite{SY1,SY2,AR,bernevig_regnault_hermanns}.

We first remind the reader of what has been termed a ``non-expandable'' basis state\cite{ortiz} in the expansion of a zero mode,
\be\label{spectral}
\ket {\psi_0} = \sum_{\{n\}}C_{\{n\}} |\{n\}\rangle\,.
\ee
Here, each $\ket{\{n\}}$ is a basis state created by a product of
single particle creation operators
$c_{i,m}^\dagger$.
We will call a basis state $|\{n\}\rangle$ in \Eq{spectral} non-expandable
if it enters the expansion with non-zero coefficient $C_{\{n\}}$ and it cannot
be obtained from any other such basis state $|\{n'\}\rangle$, also having $C_{\{n'\}}\neq 0$,
through ``inward-squeezing'' processes\cite{BH2}. That is,
\be
\ket{\{n\}}\neq c^\dagger_{l_1,j} c^\dagger_{l_2,i} c_{l_3,i-x} c_{l_4,j+x}\dotsc \ket{\{n'\}}  \,,
\ee
where a single inward squeezing process is a center-of-mass conserving inward pair hopping satisfying $i-x<i\leq j<j+x$, the $l_1...l_4$ are {\em arbitrary} Landau level indices (thus generalizing the standard notion of inward squeezing for single Landau level one-component states), and the dots represent a multiplicative string of any finite number of such inward squeezing terms.

The existence of non-expandable states in any occupancy number spectral decomposition
of the form \eqref{spectral} follows from the finiteness of the number of states
available at given angular momentum. (We may of course limit the discussion to zero modes of well-defined angular momentum without loss of generality).
It turns out, as we will show below for the present case, that such non-expandable states are subject to certain quite restrictive rules.
We will first describe the more familiar situation for single component, lowest LL states.
In this context, the rules governing non-expandable product states have been referred to as generalized Pauli principles(GPPs).\cite{BH1,BH2,BH3,ThomalePRB84:45127}
Product states satisfying these rules are generally known as dominance patterns or root patterns.
Every zero mode contains at least one non-expandable root pattern in its orbital occupancy number spectral decomposition \eqref{spectral}. Typically, a clever basis of zero modes may be chosen in a manner that there is precisely one such root pattern per zero mode. It then follows from the above that every $\ket{\{n\}}$ appearing in the zero mode's decomposition \eqref{spectral} may be obtained from its unique root pattern through inward squeezing processes. This then establishes a one-to-one correspondence between root patterns and zero modes. It is worth pointing out that
while this correspondence has been discussed for a large class of single component quantum Hall states,\cite{BH1, BH2, BH3,ThomalePRB84:45127,FlavinPRB86:125316} this was usually done by analysis of special analytic clustering conditions attributed to first-quantized zero mode wave functions. The very notion of clustering conditions  may be less clear in the presence of Landau level mixing.
Related to this,
while for single component states root patterns always represent
simple, non-entangled product states, we find it useful to relax this notion
in the multi-component or multi-Landau-level situation of interest here.
Indeed, the analysis of multi-component states\cite{SY1,SY2}
suggests the following generalization:
We will distinguish between dominance patterns and ``root states''. Dominance patterns are certain strings of symbols subject to rules we will work out below.
To each dominance pattern, we can associate a root state, which will be a fairly simple linear combination of
product states $\ket{\{n\}}$, but one possibly featuring some local entanglement. It will then follow from the rules below that the non-expandable Slater-determinants $\ket{\{n\}}$ appearing in any zero mode must appear as linear combinations of root states.
Again, a clever basis of zero modes can be chosen, where each zero mode is associated to exactly one dominance pattern, or one root state. This does, however, no longer imply that the zero mode features just a single non-expandable Slater determinant in its expansion \eqref{spectral}.


We note again that ``entangled root states'' as described above have appeared earlier in the context of multicomponent quantum Hall states.\cite{SY1,SY2}
In this context, other approaches to defining dominance patterns have been brought forth as well.\cite{bernevig_regnault_hermanns}
The approach taken here is such that, while no reference to a ``thin torus'' like geometry is made, our definition of a root state will  necessarily agree with that based on the thin torus limit.
The thin torus approach has been explored for the multi-component states discussed in Refs. \onlinecite{SY1,SY2} using first quantized analytic wave functions.
In the following, however, we argue that a
more efficient and general approach to studying the structure of root states is to forgo first quantized wave functions, and work with a second quantized form of the zero mode condition as in \Eq{zmc}. We find this particularly true in problems where degrees of freedom beyond pure guiding centers are present, e.g. spin and/or Landau level degrees of freedom. To this end we generalize the method introduced in Ref. \onlinecite{ortiz} for single Landau level, single component states to states living in multiple Landau levels.

In the following, we will write second quantized wave functions in terms of a string of numbers, e.g., 10x0x0!10..., where ! stands for an occupied orbital in the lowest LL, 1 represents an occupied orbital in the first excited LL, x represents a particle in any of the two LLs (and possibly different LLs for different occurrences of x) and 0 stands for an unoccupied orbital. Here, orbitals are arranged in the order of ascending angular momenta stating with $-1$. Before proceeding to our main results, we will state and prove a few lemmas. For definiteness, we find it useful to refer to any non-expandable Slater determinant $\ket{\{n\}}$ appearing in a zero mode as a ``root pattern''. The root state of the zero mode is then the state obtained by keeping only root patterns in \Eq{spectral}. A basis for all possible root states can then be labeled by certain dominance patterns (formal strings of symbols), as we will see below.

{\textit{Lemma 1}} There is no 101 in root patterns of any zero mode $\ket{\psi_0}$.

{\textit{Proof.}} We will use the method of contradiction and the property that any root pattern is, by definition,  non-expandable. Now let us assume that a root pattern $\ket{\{n_{\text{root}}\}}$ contains the string 101 in which 0 has angular momentum $j$. Then $\ket{\{n_{\text{root}}\}}$ can be written as $\ket{\{n_{\text{root}}\}}=c^\dagger_{1,j+1}c^\dagger_{1,j-1}\ket{\{n'\}}$. For $|x|>1$, $c^\dagger_{1,j+x}c^\dagger_{1,j-x}\ket{\{n'\}}$ must have zero coefficient in the spectral decomposition of $\ket{\psi_0}$, i.e., $\bra{\{n'\}}c_{1,j-x}c_{1,j+x}\ket{\psi_0}=0$ for $|x|>1$, otherwise $\ket{\{n_{\text{root}}\}}$ would be expandable. Thus, keeping only the $x=\pm1$ terms, $\bra{\{n'\}}Q_j^{(4)}\ket{\psi_0}=-2^{1/2-j}\sqrt{\binom{2j+2}{j+2}/(2j+2) }\langle{\{n_{\text{root}}\}} \ket{\psi_0}$, which is non-zero. This, however, contradicts the zero mode condition Eq. \eqref{q4}. Thus, 101 must be excluded from any root pattern. \hfill $\blacksquare$

Using precisely the same logic, and the respectively appropriate zero mode condition, we may further obtain the following 2 lemmas:

{\textit{Lemma 2}} There is no 11 in root patterns of the zero mode.

{\textit{Lemma 3}}
A root pattern cannot feature any simultaneous occupancy of both lowest and first excited Landau level orbitals of given angular momentum $j\geq 0$.


We then have the following stronger version of {\em Lemma 2}:

{\textit{Lemma 4}} There is no xx in root patterns of any zero mode
$\ket{\psi_0}$.

{\textit{Proof.}} According to {\textit{Lemma 2}}, there is no 11 in any root pattern, so possible configurations of xx are !!, !1 and 1!. Thus we consider $\ket {\psi_0}=(\gamma_{0,0}c^\dagger_{0,j}c^\dagger_{0,j+1}+\gamma_{0,1}c^\dagger_{0,j}c^\dagger_{1,j+1}+\gamma_{1,0}c^\dagger_{0,j+1}c^\dagger_{1,j})\ket{\{n'\}} +$ orthogonal terms where the first three terms are root patterns. As in the above, Eq.\eqref{q1} and Eq.\eqref{q3} then lead to $\sqrt{j+1}\gamma_{0,1}+\sqrt{j+2}\gamma_{1,0}=0$ and $-\sqrt{j+1}\gamma_{0,1}+\sqrt{j+2}\gamma_{1,0}=0$, respectively. Thus both $\gamma_{0,1}$ and $\gamma_{1,0}$ are zero. We then use Eq.\eqref{q256} to find that $\gamma_{0,0}$ is also zero. \hfill $\blacksquare$

The following Lemma states that x0x is allowed in root patterns, but requires local entanglement between the x-sites of the resulting root state:

{\textit{Lemma 5}} If x0x appears in root patterns of a zero mode $\ket{\psi_0}$, then the proportions of coefficients of root patterns having !0!, !01, and 10! with all other occupancies the same are $2: \sqrt{j+2}:-\sqrt{j}$, where $j$ is the angular momentum of the ``0'' in x0x.

{\textit{Proof.}}
We can write $\ket {\psi_0}=(\alpha_{0,0}c^\dagger_{0,j-1}c^\dagger_{0,j+1}+\alpha_{0,1}c^\dagger_{0,j-1}c^\dagger_{1,j+1}+\alpha_{1,0}c^\dagger_{1,j-1}c^\dagger_{0,j+1}+\beta_{0,1}c^\dagger_{0,j}c^\dagger_{1,j})\ket{\{n'\}} +$ orthogonal terms.
In the latter expression, the first three terms define three x0x root patterns related as in the statement of the lemma, whereas the fourth term is inward squeezed from these root patterns. Note that 101 must be absent in root patterns because of {\textit{Lemma 1.}}  Using Eqs.\eqref{q1}, \eqref{q3} and \eqref{q256} in a manner analogous to the proofs of the preceding lemmas, we find that $\alpha_{1,0}=-\alpha_{0,1}\sqrt{j}/\sqrt{j+2}$, $\beta_{0,1}=-2\alpha_{0,1}\sqrt{j}/\sqrt{j+2}$ and  $\alpha_{0,1}=\alpha_{0,0}\sqrt{j+2}/2$. \hfill $\blacksquare$

The next lemma involves three particles at a time. Such rules are known from single component states only in the case of 3-body Hamiltonians, but can arise here because of root state entanglement:

{\textit{Lemma 6}} There is no x0x0x in root patterns of a zero mode $\ket{\psi_0}$.

{\textit{Proof.}} From the first four Lemmas, the only allowed x0x0x in root patterns are 10!01, 10!0!, !010!, !0!01 and !0!0!. If we assume that the angular momentum of the first orbital in the above patterns is $j$, then from {\textit{Lemma 5.}}, the proportions of the coefficients of 10!0!, 10!01 and 1010! are $2: \sqrt{j+4}:-\sqrt{j+2}. $ 1010! is excluded from root patterns by virtue of {\textit{Lemma 1}}, therefore 10!0! and  10!01 are also excluded. Using the same trick, remaining three possible configurations are excluded form root patterns as well. \hfill $\blacksquare$

The last Lemma will be proven later:

{\textit{Lemma 7}}
There are no constraints on the occurrence of
x00x is in root patterns, that is, !00!, !001, 100! and 1001, and likewise for more than two zeros between occupied orbitals.

Lemma 7 is listed here for completeness, as together with the remaining lemmas,
it gives a complete set of rules for the construction of root states in one-to-one correspondence with the zero modes of the Hamiltonian. That all the root states
allowed by these rules do indeed correspond to a zero mode follows only from explicit construction of such zero modes, and will be discussed below.
The constraints imposed by Lemmas 1-6, on the other hand, can then be used to
rigorously imply that the set of zero modes thus constructed is complete.
It may be instructive, though, to see why the logic used to derive Lemmas 1-6 does not give additional constraints in the situation relevant to Lemma 7.
To briefly show this, we may write $\ket {\psi_0}=(a c^\dagger_{0,j}c^\dagger_{0,j+3}+b c^\dagger_{0,j}c^\dagger_{1,j+3}+d c^\dagger_{1,j}c^\dagger_{0,j+3}+e c^\dagger_{1,j}c^\dagger_{1,j+3}+f c^\dagger_{0,j+1}c^\dagger_{0,j+2}+g c^\dagger_{0,j+1}c^\dagger_{1,j+2}+h c^\dagger_{1,j+1}c^\dagger_{0,j+2}+i c^\dagger_{1,j+1}c^\dagger_{1,j+2})\ket{\{n'\}} +$ orthogonal terms as in the proofs of Lemmas 4 and 5. {\textit{Lemma 7}} is then related to the fact that there are eight unknown coefficients and four zero mode conditions \eqref{zero}.

We may now make precise the notion of a dominance pattern. Any root pattern satisfying Lemmas 1-4 and 6 defines a formal string of symbols ``0'', ``1'' and ``!'' as discussed above. The first character in such a string cannot be !, and the Lemmas translate into the requirements that
any 1 and any ! in such a string may have no nearest and at most one next nearest neighbor other than 0, and 101 is further disallowed. If, in all possible such strings, we send any occurrence of 10!, !01, and !0! to x0x, we will call the resulting set of strings the dominance patterns consistent with Lemmas 1-6. Examples are shown in Table \ref{table:1}.
Alternatively, we can characterize the set of all possible dominance patterns as all possible concatenations of the strings 0, 100, !00, and x0x00, with the leading character not being !.
We will refer to these concatenation rules as the GPP for dominance patterns, though this may be a slight abuse of terminology, as dominance patterns are not  generally in one-to-one correspondence with product states.
However, we may identify dominance patterns with certain states in the Fock space, consisting of the unique (up an to overall factor)
linear combination of all root patterns associated to it that also satisfies {\em Lemma 5}.
Lemmas 1-6 can then be summarized as saying that any root state of a zero mode must be a linear combination of states obtained from dominance patterns via this identification. Since the identification yields states of well-defined particle number $N$ and angular momentum $L$, we can obviously assign quantum numbers $N$ and $L$ to any dominance pattern.

Using these notions, we are able to arrive at the following important theorem(s) about the zero mode counting of the Hamiltonian \eqref{H}, where in the following, we will always imply the case $n=2$ and disk geometry:

{\textit{Theorem 1} }
At given particle number $N$ and given angular momentum $L$, the number of linearly independent zero
modes of the Hamiltonian \eqref{H}
 is no greater than the number of dominance patterns satisfying the GPP.

{\textit{Proof.}}
Assume that the number of linearly independent zero modes is greater than the number dominance patterns satisfying the GPP. Then it is possible to make a non-trivial linear combination $\ket{\psi_0}$ of such zero modes
that is orthogonal to all states
identified with these dominance patterns. Hence $P\ket{\psi_0}=0$,
where $P$ is the orthogonal projection onto the subspace spanned by all states associated to dominance patterns. On the other hand, since $\ket{\psi_0}$ is a zero mode, the definition of a root state and the lemmas imply $\ket{\psi_0}=\ket{\sf root}+\ket{\sf rest}$ where $\ket{\sf root}$ is non-zero, $P\ket{\sf root}=\ket{\sf root}$, and $\braket{\sf root|rest}=0$ .
This contradicts $\braket{{\sf root}|P|\psi_0}=0$.
\hfill $\blacksquare$

As a result, we immediately have the following

{\textit{Corollary 1.1}}
For given particle number $N$, there exist no zero modes
of the Hamiltonian \eqref{H} at angular momentum $L<L_e(N):=5/4N^2-2N$ for $N$ even, and at angular momentum
$L<L_o(N):=5/4(N-1)^2+1/2(N-3)$ for $N$ odd. If a zero mode exists at $L=L_o(N)$, it is unique, whereas for $N$ even, a zero mode at $L=L_e(N)$
can be at most doubly degenerate.

{\textit{Proof.}} The densest possible dominance patterns consistent with the GPP are, respectively, 100x0x00x0x...00x0x for $N$ odd, and
100x0x00x0x...00x0x001, 100x0x00x0x...00x0x00! for $N$ even (see also Fig. \ref{CFpattern}),
where ``densest'' means in particular that no consistent dominance patterns exist at smaller
angular momenta than the ones corresponding to these patterns, which can be seen to be $L_e(N)$ for even $N$ and $L_o(N)$ for odd $N$. Hence the statement is a special case of Theorem 1.

\hfill $\blacksquare$

For any zero mode, let $l_{\sf max}$ be the highest angular momentum among the
single particle orbitals that are at least partially occupied in that zero mode,
i.e., that have $\langle \sum_i c^\dagger_{i,l}c_{i,l}\rangle\neq 0$.
Then we finally have

{\textit{Corollary 1.2}}
Any zero mode of the Hamiltonian \eqref{H}
has $l_{\sf max}\geq 5(N-1)/2-1 $ for $N$ odd, and $l_{\sf max}\geq 5N/2-3$ for $N$ even.
Any zero modes satisfying these bounds have angular momentum
$L_o(N)$ or $L_e(N)$, respectively, and in particular the
statements about degeneracy from Corollary 1.1 apply.

{\textit{Proof.}}
Any $\ket{\{n\}}$ appearing in a zero mode either appears in its root state or can be obtained via inward squeezing from some other
Slater determinants appearing in the root state. Hence the $l_{\sf max}$
of the zero mode is the same as that of its root state, which in turn is the highest occupied orbital among dominance patterns contributing to the root state. For given $N$, the dominance patterns of smallest $l_{\sf max}$ are those referenced in the proof of Corollary 1.1, and these have the  $l_{\sf max}$ values given in the statement of Corollary 1.2, which hence follows.
\hfill $\blacksquare$

If we define the filling factor $\nu$ of a zero mode as $N/l_{\sf max}$,
then Corollary 1.2 implies that the densest (highest) filling factor for which
zero modes exist is bounded from above by $2/5$ in the thermodynamic limit.
This bound is, of course, saturated, as the corresponding wave function is known.\cite{Jain, ReMac}
So far, the statements derived here constitute upper bounds on the number of zero modes of the Hamiltonian \eqref{H}.
In the following, we will be concerned with the question
whether these bounds are saturated,
and how the resulting zero mode counting is related to the mode counting in the effective edge theory.
\setlength\extrarowheight{2pt}
\begin{table}[ht]
\begin{tabular}{l|@{\hspace{2em}} l}
 \hline
\hline

\hline
a)&100x0x00x0x00x0x00x0x\\
b)&100x0x00x0x00x0x001001\\
c)&100x0x00x0x00x0x00!00!\\
d)&100x0x00x0x00x0x00!001\\
e)&100x0x00x0x00x0x00100!\\
f)&100x0x00x0x00x0x000x0x\\
g)&100x0x00x0x00x0x0010001\\
h)&100x0x00x0x00x0x001000!\\
i)&100x0x00x0x00x0x00!000!\\
j)&100x0x00x0x00x0x00!000!\\
k)&100x0x00x0x00100x0x001\\
l)&100x0x00x0x00100x0x00!\\
m)&100x0x00x0x00!00x0x001\\
n)&100x0x00x0x00!00x0x00!\\
o)&100100x0x0000!00x0x0001001\\
\hline
 \end{tabular}
\caption{
Some dominance patterns consistent with Lemmas 1-6 for $N=9$ particles. The leading position corresponds to single particle angular momentum $L_z=-1$ and can only be 0 (empty) or 1 (first excited Landau level).
a) Unique dominance pattern at smallest angular momentum $L=83$. b)-e) All consistent patterns with $\Delta L=1$ relative
to the ground state. f)-n) All consistent patterns  with $\Delta L=2$. o) A consistent pattern with higher $\Delta L=19$.
As is shown in the text, the number of consistent patterns at given $\Delta L$ equals the dimension of the
zero mode subspace of the $n=2$ Hamiltonian \Eq{H}.
}
\label{table:1}
\end{table}

\section{Zero mode counting and edge theory\label{countedge}}

As argued in the introduction,
the zero mode condition derived from a good quantum Hall parent Hamiltonian will
 not only characterize the incompressible quantum fluid sufficiently uniquely, but also encode the proper edge theory of the system.
The rules derived in the preceding section thus far only suggest a certain zero mode structure, but, with the exception of (the yet unproven) Lemma 7, only constrain this structure without guaranteeing the existence of any zero modes.
It is, however, worth noting that all of this was derived from the second quantized operators $Q^{(i)}_R$ alone, and, if we took Lemma 7 for granted, the entire zero mode structure in terms of dominance patterns would follow correctly from this analysis.
To prove Lemma 7 and thus establish the complete zero mode structure of \Eq{H} with $n=2$,
we briefly make contact with the first quantized presentation of zero modes,
though at least in part we will see below that an operator-based approach
could also be envisioned. (In all aspects, such an operator-based approach has been constructed by some of us previously for the $n=1$ case related to the $1/3$-Laughlin state, and in fact for all the Laughlin states.\cite{ortiz,Chen14,Mazaheri14} We will comment more on the situation below.)

The analysis of Sec. \ref{2ndq} implies that a sufficient (and necessary) property
of any zero mode is that the associated analytic many-body wave function contains the factor $(z_i-z_j)^2$ for all $i,j$. This is, in fact, a quite special property of the cases $n=1$ and $n=2$ of \Eq{H}. More generally, zero modes of \Eq{H} may be linear combinations of terms containing the factors $(z_i-z_j)^2$, $(z_i-z_j)(\bar z_i-\bar z_j)$, and $(\bar z_i-\bar z_j)^2$, which, by symmetry, must be true for all $i,j$.
That is, a zero mode vanishes at least to second order in the separation of any pair of coordinates. For $n\leq 2$, however, the third term is prohibited by Landau-level projection, and the second then always necessitates another factor of $z_i-z_j$ by anti-symmetry, such that the first term still covers all possible cases for having a second order zero. This renders the $n=2$ of \Eq{H} rather special. While the presence of the first excited Landau level allows terms in $\bar z_i$ to be present in the wave function, the zero mode condition can thus be stated only in terms of the holomorphic variables $z_i$. Indeed, it is only for $n\leq 2$ that the ground state of \Eq{H} is in the Jain sequence of states.\cite{wu2016new}

Thanks to the work done in the preceding section, for now it will do to note that
divisibility of the wave function by $\psi_{1/2}=\prod_{i<j} (z_i-z_j)^2$, the bosonic $\nu=1/2$ Laughlin-Jastrow factor, is a sufficient criterion for a wave function to be a zero mode. The necessity of this criterion (for $n=2$), i.e., the completeness of the resulting zero mode space,
can then alternatively be inferred
from Theorem 1. This route will set the stage for the larger $n$ Hamiltonians as well (where we are currently not aware of any alternative).
As an added benefit, this will establish the one-to-one correspondence between dominance patterns satisfying the rules given above and zero modes of the Hamiltonian.

We thus consider zero mode wave functions of the form $\psi_{1/2} p(z_1,\bar z_1,\dotsc ,z_N,\bar z_N)$, where $p$ is an arbitrary polynomial of the requisite
anti-symmetry and at most first order in the $\bar z_i$ (so as for $\psi_{1/2} \,p$ to be contained within the first two Landau levels), and we drop the obligatory Gaussian factor for simplicity. It is clear that a suitable basis for these polynomials is given by
$S_{\{\mathfrak{n}\}}(z_1,\bar z_1,\dotsc)$,  where $S_{\{\mathfrak{n}\}}$ is a Slater determinant of single particle states in the lowest and first excited Landau level, with occupancies determined by a set of occupancy numbers $\{\mathfrak{n}\}$. \footnote{If there were any doubts as to the completeness of these Slater determinants for present purposes, this would follow below from the fact that all possible dominance patterns are obtained in this way.}
Hence we wish to study zero modes of the form
\be\label{1qzm}
\psi_{1/2}(z_1,\dotsc) S_{\{\mathfrak{n}\}}(z_1,\bar z_1,\dotsc)\,.
\ee
We note that zero modes of this form are naturally viewed as composite fermion (CF) states, where any fermion forms a composite object with two flux quanta. In particular, if the CF-occupancy configuration $\{\mathfrak{n}\}$ is chosen to represent two equally filled Landau levels, one recovers the Jain-2/5 state, and one easily verifies that this state saturates the bounds of the Corollaries of the last section. Therefore, the Jain-2/5 state is the densest zero mode of \Eq{H} for $n=2$, unique up to the twofold degeneracy mentioned in Corollary 1.1 (see below).

\begin{figure}
\centering
{\includegraphics[width=7cm]{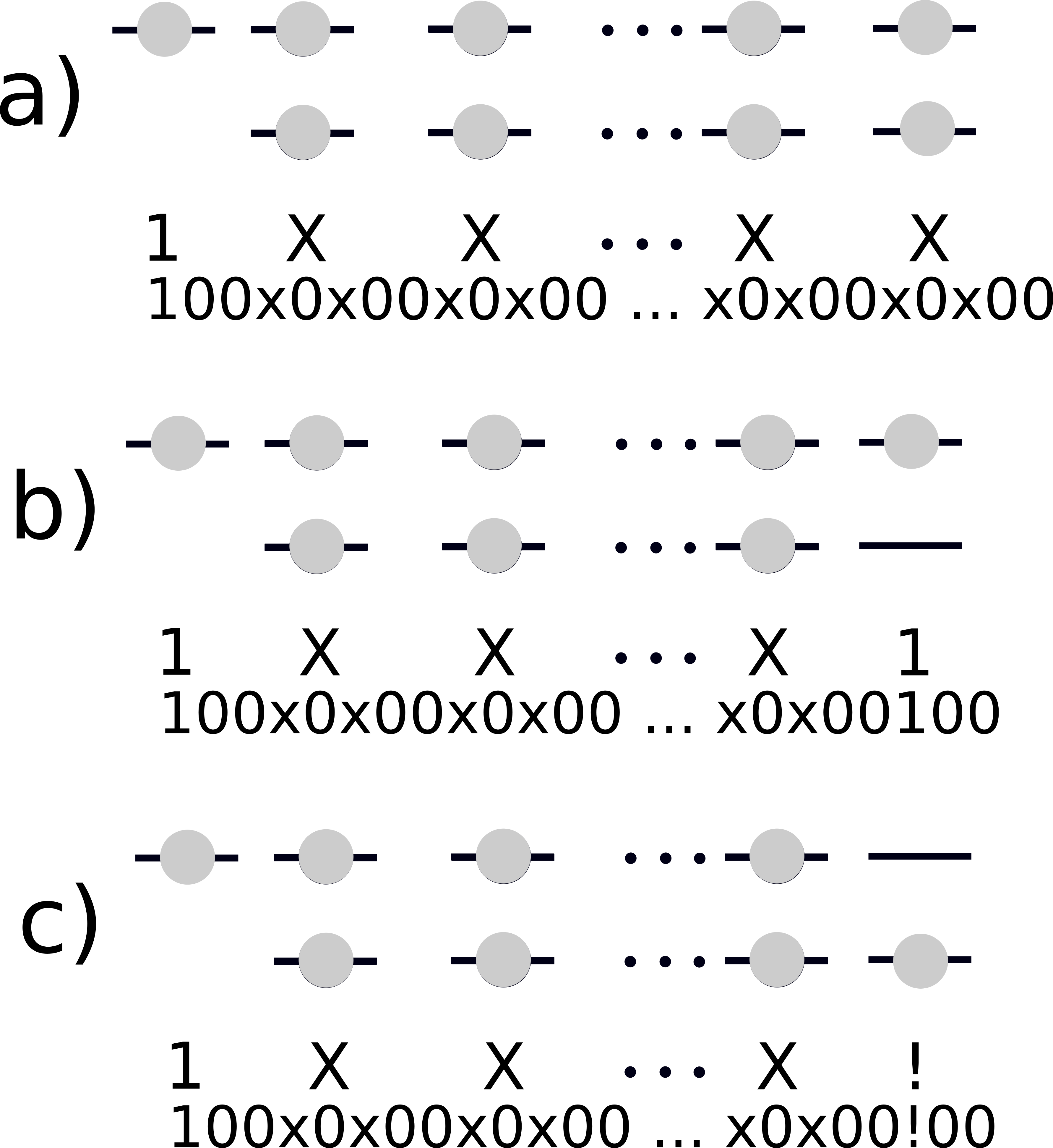}}\vspace{2mm}
\caption{Composite fermion occupancy patterns and resulting dominance patterns. Three different cases are shown. Level diagrams show composite fermion occupancies, followed by a more symbolic composite fermion occupancy pattern and
the associated dominance pattern as explained in text. a) corresponds to the densest (minimum angular momentum) zero mode for odd particle number, followed by the two configurations corresponding to the doubly degenerate densest zero modes for even particle number (b) and c)). Note that only the dominance patterns manifestly encode the total angular momentum of the state. More general dominance patterns consistent with Lemmas 1-6, and thus in one-to-one correspondence with zero modes (see text), are shown in Table \ref{table:1}.
}
\label{CFpattern}
\end{figure}

We emphasize that while notationally similar to the {\em electron} occupancy numbers $\{n\}$ labeling basis states
in \Eq{spectral}, the labels $\{\mathfrak{n}\}$ represent {\em composite fermion} occupancy numbers and must be well distinguished from the labels $\{n\}$.
To analyze the dominance patterns underlying the zero modes \eqref{1qzm}, we make use of well known rules\cite{regnault} for products of polynomials with known root patterns, generalized to the case where non-holomorphic variables (or more than a single Landau level) are present. Every CF-Slater determinant configuration $S_{\{\mathfrak{n}\}}(z_1,\bar z_1,\dotsc)$ is naturally
its own root state, as it is the only Slater determinant appearing in its wave function. The associated CF-occupancy pattern $\{\mathfrak{n}\}$ may now be thought of as a string made up of characters $X$, $0$, $1$, and $!$. The last three characters have the analogous meaning as in our notation for root patterns of full zero mode wave functions (but refer to CFs), and $X$ now means a double occupancy of the associated angular momentum state in both Landau levels.
As before, the first character can only be $1$ or $0$, see Fig.\ref{CFpattern}.
Moreover, as is well known,\cite{RH} the bosonic Laughlin factor $\psi_{1/2}$ has a root state given by the pattern $!0!0!0!0\dotsc$.
Dominance patterns may generally be associated to partitions $l_N+l_{N-1}+\dotsc+l_1=L$, where
$l_i\geq l_{i+1}$ is the angular momentum of the $i$th particle in the pattern, and $L$ is the total angular momentum of the pattern.
When two wave functions whose root states have dominance patterns with partitions $\{l_i\}$ and $\{l_i'\}$, respectively, are multiplied, the resulting wave function has a root state whose dominance pattern has the partition $\{l_i+l_i'\}$. It is easy to see that these rules, when applied to the present situation, imply that the multiplication of $\psi_{1/2}$ by the Slater determinant $S_{\{\mathfrak{n}\}}$ leads to a wave function with a dominance pattern obtained from the pattern associated to $\{\mathfrak{n}\}$
as follows. The character $!$ is replaced with $!00$, ($!\rightarrow !00$, rule 1). An $X$ in the CF-pattern
corresponds to the case where $l_i=l_{i+1}$ in the associated partition, signifying two particles with identical angular momenta but different Landau level indices. The resulting ambiguity in ordering these two particles leads to the situation described as x0x in the dominance pattern of the resulting zero mode, i.e., we have the rule $X\rightarrow $x0x00 (rule 2).
That the underlying configurations $!0!$, $10!$, and $!01$ indeed occur with the ratios claimed by Lemma 5 could be verified directly from \Eq{1qzm}, but this is not necessary, since \Eq{1qzm} is definitely a zero mode, and then the proof of Lemma 5 applies.
A ``$1$'' in the CF-pattern associated to $S_{\{\mathfrak{n}\}}$ leads to at least two root patterns in the root state of \Eq{1qzm}, one obtained from the replacement $1\rightarrow 100$ (rule 3.a), and one from $1\rightarrow !00$ (rule 3.b). 
However, it is clear that if we ignore rule 3.b for the moment, rules 1-3.a establish a one-to-one correspondence (see Fig.\ref{CFpattern}) between CF-occupation number patterns $\{\mathfrak{n}\}$ of $N$ particles occupying orbitals with angular momentum up to $l_{\sf max}$ and permissible dominance patterns of $N$ particles occupying orbitals with angular momentum up to $l_{\sf max}+2(N-1)$ (where the addition of $2(N-1)$ can be thought of as being due to flux attachment.) Let us now denote
a dominance pattern
satisfying the GPP of the preceding section by $p$ and the associated root state  by $|p\rangle$. Let us choose an ordering of these patterns such that the number of $1$s in the pattern increases monotonously for patterns associated to the same partition $\{l_i\}$.
Furthermore, we may order patterns associated to different partitions according to increasing $S(\{l_i\}):=\sum_i l_i^2$.
Finally, let us order the CF-occupancy patterns $\{\mathfrak{n}\}$ in the same way, by means of the one-to-one correspondence. We then see
that the matrix
\be\label{overlap}
  C_{p,\{\mathfrak{n}\}}=\langle p|\psi_{1/2}S_{\{\mathfrak{n}\}}\rangle
\ee
is upper triangular\footnote{For, let $p_{\mathfrak{n}}$ be the pattern that is associated to ${\mathfrak{n}}$. Then by design, any $p'$ different from $p_{\mathfrak{n}}$ but having the same partition $\{l_i\}$ must come before $p_{\mathfrak{n}}$ in order for the overlap (\ref{overlap}) to be non-zero.  Likewise, any such $p'$ corresponding to a different partition $\{l'_i\}$ would be obtainable from the dominant pattern $p_{\mathfrak{n}}$ via inward squeezing, and thus have smaller $S(\{l'_i\})$.} with non-zero diagonal and thus invertible.
This implies that for each dominance pattern $p$ satisfying the GPP, there is a superposition of zero modes of the form \eqref{1qzm} that is dominated precisely by the associated root state $\ket{p}$ , with no other of the states $\ket{p'}$ present in its spectral decomposition \eqref{spectral}. This establishes thus the one-to-one correspondence between zero modes and dominance patterns satisfying the GPP.

We will now discuss that the counting of zero modes at a given angular momentum and particle number 
that follows from the construction of dominance patterns above agrees with counting of edge states in the effective edge theory.
We will argue that there is a weaker and a stronger version of this statement. The weaker version, often found in the literature, is concerned with the number of zero modes/edge modes ${\cal N}(\Delta L)$, where $\Delta L$ is the angular momentum relative to the ground state at fixed particle number. In the thermodynamic limit of large particle number $N$, this quantity is not expected to depend (much) on $N$. We will see that the counting problem defined by ${\cal N}(\Delta L)$ can be conveniently addressed in terms of CF-patterns. However, the quantity
${\cal N}(\Delta L)$ is not sensitive to all aspects of the $K$-matrix describing the edge theory.
Indeed, the $K$-matrix of any Jain state is congruent to a matrix of the form $K'=W^T K W= m J_n+1\!\!1$,
where $J_n$ is an $n\times n$ matrix of ones, and $W$ is an $SL(n,\mathbb{Z})$ matrix.
$K'$ has precisely one eigenvalue different from $1$, which is non-degenerate with eigenvector $t$ describing charged excitations.
The quantity ${\cal N}(\Delta L)$
is only sensitive to neutral excitations orthogonal to $t$, which always lie in the eigenvalue $1$ eigenspace of $K'$. In particular, ${\cal N}(\Delta L)$ does not distinguish
between Jain states that have the same number of edge branches. (For example, ${\cal N}(\Delta L)$ does not distinguish different Laughlin states; see, e.g., the discussion in Ref. \onlinecite{wenbook}.)
In contrast, we may consider the number of zero modes ${\cal N}(N,L)$ at given particle number and given {\em total} angular momentum, which, among other things, also keeps track in absolute terms of how angular momentum changes with particle number. We will show that this quantity, when evaluated for the present microscopic Hamiltonian, captures all aspects of the $K$-matrix of the edge theory.

To make things concrete, we consider the edge theory of the Jain-$2/5$ states in the form\cite{wen_toprev}
\begin{equation}\label{Hedge}
H=\frac{1}{4\pi}\int dx\, V_{ij}:\partial_x\phi_i\partial_x\phi_j: - \frac{\mu_i}{2\pi} \int dx \,\partial_x \phi_i\,,
\end{equation}
where $i,j=1,2$ describe two bosonic edge modes through phase fields $\phi_i(x)$ and associated densities
$\rho_i=\frac{1}{2\pi}\partial_x \phi_i$, satisfying the Kac-Moody algebra $[\rho_i(x),\rho_j(x')]=(K^{-1})_{ij}\frac{i}{2\pi}\partial_{x'}\delta(x-x')$. The colons imply normal ordering with respect to finite momentum modes defined below.
$K_{ij}$ is a characteristic matrix that together with the charge vector $t_i$ defines the edge theory. The Jain- or hierarchy-2/5 edge
can be described by $K=\left ( \begin{smallmatrix} 3 & 2 \\ 2 & 3 \end{smallmatrix} \right )$ and $t=(1,1)$,
where $t$ is defined such that $\rho_e=\sum_i t_i\rho_i$ represents the physical electron charge. In the following, we will pay special
attention to the zero momentum modes of the
densities $\rho_i$, which we will write as
$N_i/(2\pi R)$, where $R$ is the radius of the
quantum Hall fluid. Physical operators
must respect the integer character of the $N_i$.\cite{wen_toprev}
We note in passing that close formal relations\cite{MR} between the edge theory conformal blocks and CF wave functions have been explored in detail in Ref. \onlinecite{HanssonPRB76:075347}.

\Eq{Hedge} describes an edge with general interaction matrix $V_{ij}$ between densities and with general chemical potentials $\mu_i$ coupling to the integer charges $N_i$.
The latter control both the total particle number as well as the radial spatial separation between the two edge branches, which, in the limit of large separation, define two individual edges between a 2/5-phase and a 1/3-(Laughlin-)phase and between a 1/3-phase and vacuum, respectively. On general grounds,\cite{Read_edgetheory} a close relation is expected between the spectrum of the edge Hamiltonian and the angular momentum operator of the fluid, if the interactions are so tuned that the edge theory is conformally invariant. This requires all edge modes to travel with the same velocity $v$. Is is easy to see that this can be achieved by letting $V_{ij}=vK_{ij}$, leading to the equation of motion $\partial_t \rho_i+v\partial_x \rho_i=0$. With this, we then look at the mode expansion of \Eq{Hedge}:
\be \label{edgemode}
\begin{split}
H=\frac{v}{2R}(3N_0^2+3N_1^2+4N_0N_1)-\mu_0N_0-\mu_1N_1+\frac{v}{R}\, P,\\
P=\sum_{j=0,1}\sum_{n>0} n\, b^\dagger_{j,n} b_{j,n}\,.
\end{split}
\ee

Here, the $b_{j,n}^\dagger$ ($b_{j,n}$)
are appropriate linear combinations of the
positive (negative) Fourier components of the
$\rho_i(x)$ satisfying $[b_{j,n},b_{j',n'}^\dagger]=\delta_{j,j'}\delta_{n,n'}$, $n=1,2,\dotsc$.

For the purpose of comparing the dimensions of zero mode spaces and edge mode spaces for various sectors, it is useful to identify the quantum numbers $N_0$, $N_1$ of the edge theory with the CF-numbers in the lowest and first excited LL, respectively, in zero modes of the form \eqref{1qzm}.
We first appeal to the one-to-one correspondence between CF-occupancy patterns of fixed $N_i$ and excitations of the edge theory, likewise for fixed $N_i$.
This is a standard result in bosonization,\cite{stone1994bosonization} applied here to the case of two chiral branches.
Let us denote the CF-state with
``densest'' (minimum angular momentum) CF-occupancy pattern for given $N_i$ by $\ket{N_0,N_1}_{\sf CF}$.
Then the one-to-one correspondence
between CF-states and edge states at fixed $N_i$ applies to all CF-states
whose angular momentum relative to
 $\ket{N_0,N_1}_{\sf CF}$ is smaller than a cutoff given by particle number: $\Delta L\lesssim N_i$ (c.f., e.g., Ref. \onlinecite{Mazaheri14}). That is, the number of such CF zero modes of given $N_i$ and $\Delta L$ relative to
$\ket{N_0,N_1}_{\sf CF}$ is equal to the number of edge states described by \Eq{edgemode} of fixed $N_i$ and ``edge momentum'' $P=\Delta L$.

We note, however, that counting at fixed $N_i$ is an artificial constraint from the point of view of the microscopic theory, as these quantum numbers do not correspond to any local (or even Hermitian) conserved quantities in the microscopic theory. Moreover, counting subject to this constraint contains no information about the $K$-matrix (except for its dimension). To make a statement that is both more physical and stronger, we now claim that for proper choice of chemical potentials $\mu_i$ and up to a scale factor $v/R$ we will let equal to $1$, for any given particle number $N=N_0+N_1$,
the degeneracies of the eigenvalues of the angular momentum operator of the macroscopic theory, projected onto the zero mode subspace of \Eq{H}, are exactly the same as the degeneracies of the energy eigenvalues
of the edge Hamiltonian \Eq{edgemode}. That is, the number ${\cal N}(N,L)$ introduced above for the microscopic Hamiltonian is identical to the degeneracy of the energy $E=L$ of \Eq{edgemode} for given $N=N_0+N_1$. Loosely speaking, the edge Hamiltonian \Eq{edgemode} {\em is} the zero-mode-projected angular momentum operator of the microscopic theory.

It is sufficient to show that edge states with $P=0$ and given $N=N_0+N_1$ have an energy equal
to the angular momentum of the CF ``vacua'' $\ket{N_0,N_1}_{\sf CF}$ defined above. For then, it follows that all states identified within each $N_0$, $N_1$ sector via bosonization must also have identical eigenvalues for, respectively, energy (in \Eq{edgemode}) and angular momentum (in the microscopic theory). The choice of $\mu_i$ for which this is true
 is totally determined by the requirement that $N_0=N_1=1$ leads to angular momentum $L=1$ in the microscopic theory, whereas $N_0=0$, $N_1=1$ leads to $L=-1$, giving $\mu_0=3/2$, $\mu_1=5/2$ in \Eq{edgemode} ($v/R=1$, $P=0$). It thus suffices to show that the minimum angular momentum states $\ket{N_0,N_1}_{\sf CF}$ have $L$ equal to
\be \label{Lmin}
  L_{\sf min}= \frac{3}{2}(N_0+N_1)(N_0+N_1-1)-N_1(N_0+1)\,.
\ee
That this is indeed the case can easily be established, e.g., from \Eq{1qzm} or
by studying the
densest possible dominance patterns of given $N_0$, $N_1$ (examples are b) and c) in Table \ref{table:1} for $N_0=3$, $N_1=6$ and $N_0=5$, $N_1=4$, respectively).

The above establishes that the counting of microscopic zero modes at given particle number $N$ and angular momentum $L$ is exactly the same as that of energy eigenmodes in an appropriately scaled edge Hamiltonian describing the $2/5$-edge. While the counting can be done in terms of CF-patterns, as expected in any system that can be understood in terms of non-interacting CFs, we have shown that counting can be done equally well in terms of dominance patterns. In this regard, it is worth noting that CF occupancy patterns as defined above manifestly encode only {\em changes} in angular momentum at {\em fixed} particle number. Obtaining the absolute angular momentum of a CF-state described by a given CF occupancy pattern requires additional information about the number of flux quanta each composite fermion carries. In contrast, the total angular momentum of the associated (root) state is manifest in dominance patterns. The set of rules governing the composition of valid dominance patterns can thus be interpreted as a set of minimal rules to construct the quantity ${\cal N}(N,L)$ from certain local building blocks (see discussion above Theorem 1 and caption of Table \ref{table:1}). The fact that this then reproduces edge mode counting is the property that one expects a good GPP to have.
 We thus find that the present Hamiltonian does not only fully fall into the ``zero mode paradigm'' expected of special quantum Hall parent Hamiltonians, but is also linked to a GPP which facilitates the pertinent counting.
It should be clear that our arguments leading
from FQH Hamiltonians admitting zero modes to GPPs governing dominance patterns have a very general character.
If such a Hamiltonian satisfies the zero mode paradigm, the implied GPP must then reproduce edge mode counting from local rules as demonstrated above.
  We will argue below that this general connection between the existence of zero modes and GPPs imposes useful constraints on settings in which ``good'' (zero mode paradigm) parent Hamiltonians may be constructed. We caution, however, that there are modified versions of this paradigm, as, e.g., realized in the parent Hamiltonian of the anti-Pfaffian state.\cite{levinAP, leeAP}
  Here, the equivalent of zero mode counting would describe an edge with a $\nu=1$ integer quantum Hall state, as opposed to vacuum.

We note that the quantity ${\cal N}(N,L)$ is in principle robust to sufficiently weak rotationally invariant perturbations. Here, ``weak'' means sufficiently small compared to the gap separating low-energy modes from the rest of the spectrum at given $L$. Under such conditions, ${\cal N}(N,L)$ may thus even survive some degree of edge reconstruction. However, it is clear that this quantity is directly meaningful only in exceptionally clean systems. The more robust features of edge mode counting can be probed experimentally in momentum-resolved tunneling.\cite{KangNature403:59,YangPRL92:056802,HuberPRL94:016805, MelikidzeIntJModPhysB18:3521, SeidelPRB80:241309,WangB81:035318}

\section{Zero mode generators\label{zmgn}}
While results from the preceding section establish the full zero mode structure of the Jain-2/5 state parent Hamiltonian, we mention here an alternative approach more in line with our general philosophy of working with the operator algebras of the second quantized problem.
Such an approach has been carried out earlier by some of us\cite{ortiz,Chen14,Mazaheri14} for the Laughlin states and their parent Hamiltonians. One attractive feature of this approach is its resulting in a ``microscopic bosonization dictionary'', where operators present in the effective edge theory are identified with second-quantized microscopic operators that interact with the microscopic Hamiltonian in exactly the way expected from the effective theory. Another motivation to consider this route is the fact that, in the single Landau level example of Refs. \onlinecite{ortiz,Chen14,Mazaheri14}, Read's order parameter of the Laughlin state\cite{readOP} appeared naturally (in a fully second quantized form). Clearly, an analogous construction for the Jain-2/5 state would be of great interest. Here we will report some preliminary results regarding this approach, leaving details for future work.

We begin by identifying four sets of single particle ``zero mode generators'':
 \be
P_d^{(1)}=\sum_{r=-1}^{+\infty}\sqrt{\frac{(r+d)!}{(r+1)!}}c^\dagger_{0,r+d}c_{1,r}\quad d\geq 1,\ee
\be\label{P2} \begin{split}P_d^{(2)}&= \sum_{r=0}^{+\infty} \sqrt{\frac{(r+d)!}{r!}}c^\dagger_{0,r+d}c_{0,r}\\& +\sum_{r=-1}^{+\infty}\sqrt{\frac{(r+d+1)!}{(r+1)!}}c^\dagger_{1,r+d}c_{1,r} \quad d\geq 0,\end{split}\ee
\be
\begin{split} P_d^{(3)}&=\sum_{r=-1}^{+\infty} \Big((r+d+1)\sqrt{\frac{(r+d)!}{(r+1)!}}c^\dagger_{0,r+d}c_{1,r}\\ & +\sqrt{\frac{(r+d+1)!}{(r+1)!}}c^\dagger_{1,r+d}c_{1,r}\Big)\quad d\geq 0,\end{split}\ee
\be \begin{split}P_d^{(4)}&=\sum_{r=0}^{+\infty} \Big(\sqrt{\frac{(r+d+1)!}{r!}}c^\dagger_{1,r+d}c_{0,r}\\&+(r+d+1)\sqrt{\frac{(r+d)!}{r!}}c^\dagger_{0,r+d}c_{0,r}\big)\\ & -\sum_{r=-1}^{+\infty} \Big((r+1)\sqrt{\frac{(r+d+1)!}{(r+1)!}}c^\dagger_{1,r+d}c_{1,r}\\ &+(r+1)(r+d+1)\sqrt{\frac{(r+d)!}{(r+1)!}}c^\dagger_{0,r+d}c_{1,r}\Big)\quad d \geq-1.\end{split}\ee

These generalize the single set of zero mode generators identified for the $n=1$ (Laughlin-state) case earlier.\cite{ortiz,Chen14,Mazaheri14}
Their algebraic properties can be summarized as follows. Details will be published elsewhere.\cite{tobepublished} By themselves, the $P^{(i)}_d$ form a graded Lie-algebra, where the grading is furnished by the label $d$. Explicitly, this means that $[P^{(i)}_d,P^{(j)}_{d'}]$ is a linear combination of $P^{(k)}_{d+d'}$, $k=1\dotsc 4$.
This graded Lie-algebra can be extended by the $T_R^{(i)}$, or, alternatively, the operators appearing on the left hand side of \Eq{zero} defining the zero mode condition, where the grading is now provided by the label $-2R$.
While commutators between different
$T_R^{(i)}$ of course vanish,
commutators of the form
$[T_R^{(i)}, P^{(j)}_d]$ give linear combinations of $T_{R-d/2}^{(k)}$, $k=1\dotsc 4$.
This last property justifies the term ``zero mode generators''.
It assures that,  when any $P^{(i)}_d$ acts on a zero mode $\ket{\psi}$ (and does not give zero), it generates another zero mode, because all commutators $[T_R^{(i)}, P^{(j)}_d]$ vanish inside the zero mode subspace.\cite{ortiz}
Note also that $P^{(i)}_d$ increases the angular momentum of the zero mode by $d$.
It thus clear that  the  $P^{(i)}_d$
have properties that are similar to those of the  mode operators  $b_{i,d}^\dagger$ ($i=0,1$)in the effective edge theory.  This leads to the obvious question why we  found  more than two  sets of  $P^{(i)}_d$ operators.  Although we  must carefully distinguish between  electron and CF occupancy numbers, it is clear that the operator $P^{(1)}_d$ gradually  depopulates the first excited Landau level. This will also reduce the number of CFs in the first excited Landau level. Note that the operator is nilpotent (for fixed particle number): A sufficiently large power of $P^{(1)}_d$ will certainly annihilate the state. We may thus interpret  $P^{(1)}_d$ as an operator that creates edge excitations  of the kind generated by the operators  $b_{i,d}^\dagger$ in the effective edge theory, but at the same time lowers the quantum number $N_1-N_0$.  To identify zero mode operators that, like the operators $b_{i,d}^\dagger$ create  independent branches of edge excitations that do {\em not} affect  $N_1-N_0$, we must find two commuting linear combinations of the $P^{(i)}_d$ that  are {\em not} nilpotent.
These criteria  are  satisfied by
$dP^{(1)}_d+P^{(2)}_d$ and $P^{(3)}_d$.
The other two linear combinations of the  $P^{(i)}_d$ operators will correspond to  operators in the edge theory that do change the quantum number $N_1-N_0$ (or else are not independent of the former).  We  have indeed shown that $P^{(4)}_0$
can  be used to connect one of the  two degenerate lowest angular momentum zero modes at even particle number(see Sec. \ref{countedge}) to the other.\cite{tobepublished}  These considerations make it feasible that  by acting with  combinations  of products of the operators $P^{(i)}_d$ on a lowest angular momentum zero mode, 
we can generate all  zero modes at fixed particle number.  We leave this as a conjecture for future work.
Moreover, in Ref. \onlinecite{Chen14} we have succeeded in  constructing a microscopic operator that,  when acting on the smallest angular momentum zero modes in the $n=1$ (Laughlin) case,  leads to  the corresponding zero mode with the total particle number increased by 1.  This can be interpreted as a microscopic realization of the operator of the edge theory that  raises the quantity $N_0+N_1$.  It is here where the connection with the order parameter of the Laughlin state can be made. We will also leave the generalization of this operator to the present situation as an interesting problem  for future work.




\section{Second quantization on the sphere\label{sphere}}

In this section, we wish to make contact with
previous studies that seem to have focused on the sphere.\cite{ReMac,Wen1991}
One question that has been addressed by earlier works is the uniqueness of the ground state
whenever the number of flux quanta is chosen to be $2s=5/2N-{\cal S}$ where ${\cal S}=4$ is the topological shift of the Jain=2/5 state.  This requires the particle number $N$ to be even.
 We have seen above that  for even $N$ there generally is no unique ground state
in the disk geometry. However, the statement is nonetheless correct on the sphere. While earlier confirmations of this uniqueness
seem to have rested at least in part on numerics for finite particle number, the methods established above suggest several routes to establish this fact analytically. Indeed, the statement becomes immediate
once lemmas 1-6 have been translated to the sphere.
For this we will also have to briefly discuss the second quantized form of the $n=2$ Hamiltonian on the sphere, which we also believe to be of benefit for future reference.

\begin{figure}
\centering
\vspace{.5cm}
{\includegraphics[width=6cm]{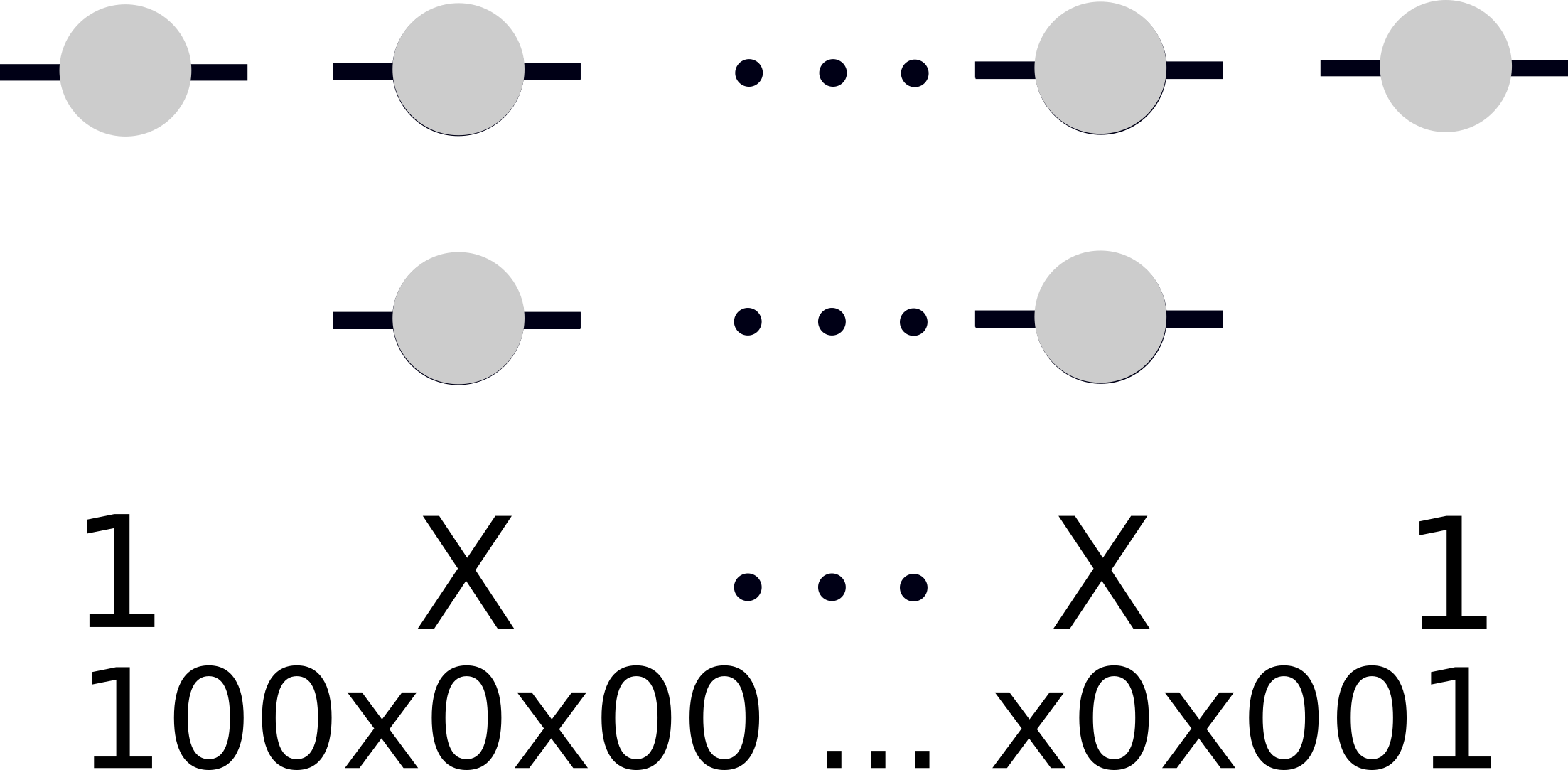}}
\caption{Same as Fig. \ref{CFpattern}, but for the sphere, where the first excited LL has one more orbital at both maximum and minimum $L_z$, for both electrons and composite fermions. Shown (bottom line) is the resulting unique dominance pattern for a sphere satisfying $2s=\frac{5}{2}N-4$, where $2s$ is the number of flux quanta penetrating the sphere. }
\label{spherepattern}
\end{figure}

We first remind the reader that a sphere threaded by $2s$ flux quanta has a Landau level
structure where the $i$th Landau level has $2(s+i)+1$ orbitals.\cite{haldane_hierarchy}
Moreover, the $i$th Landau level transforms under rotations according to the spin $s_n=s+i$ representation of SU(2). Working with eigenstates of the z-component of angular momentum,
basis states within a given Landau level thus vary from $L_z=-s-i$ to $s+i$.
Specializing to $n=2$, this means that not only the smallest possible $L_z$ is unique to the first excited Landau level (as is $L_z=-1$ in the disk geometry), but so is the largest $L_z$.
The situation is depicted in Fig.\ref{spherepattern}.  We see that boundary conditions on the left end are then  exactly the same as on the right.
When the filling factor is given by $2s=5/2N-4$,  the application of Lemmas 1-6 then leads to a unique dominance pattern. By Theorem 1, this in turn yields the uniqueness, as a zero mode, of the corresponding Jain-2/5 state  on the sphere.  Likewise,  there cannot be any zero modes for $2s<5/2N-4$, due to the impossibility to construct permissible dominance patterns under such conditions.

To  establish the above, we now turn to
the second quantized presentation of $n=2$ Hamiltonian on the sphere.
We will work with the stereographic projection of the sphere introduced in this context in Ref. \onlinecite{RRcount}:
\be\label{sp1}
z=\tan\frac{\theta}{2}e^{-i\phi}\,,
\ee
where  $\theta$ and $\phi$ are the usual polar and azimuthal angles on the sphere, respectively.
With  this, the rotationally invariant volume element on the sphere becomes
 $\sin\theta \,d\theta d\phi=\sqrt{g(z)}dzd\bar{z}$
 with $g(z)=(1+z\bar{z})^{-4}$.
 The rotationally invariant analog of \Eq{H} is then
\begin{equation}\label{sp2}
H=P_n\frac{\partial_{z_1}\partial_{\bar{z}_1}\delta(z_1-z_2)\delta(\bar{z}_1-\bar{z}_2)}{\sqrt{g(z_1)g(z_2)}}P_n\,.
\end{equation}
Moreover, using the gauge
$A=-\frac{2s}{e}\cot{\theta}\hat{e}_\phi$,
the relevant lowest and first excited Landau level  single particle states  have wave functions
\be
\begin{split}
\eta_{0,m}(z)&={\cal N}_{0,m}\,{z^{s-m}} G_0(z,\bar z),\\
\eta_{1,m}(z)&={\cal N}_{1,m}\,[(1+s+m)z\bar{z}-(1+s-m)]{z^{s-m}}G_1(z,\bar z)
\end{split}
\ee
where the normalization factors are
\[
\begin{split}
{\cal N}_{0,m}=\sqrt{{(2s+1)!}/[{(s+m)!(s-m)!}]},\\
{\cal N}_{1,m}=\sqrt{{(2s+3)!}/[{2(1+s)(1+s+m)!(1+s-m)!}]}\end{split}\]
and furthermore $G_n(z,\bar z)={\bar{z}^{s/2}}/{[z^{s/2}(1+z\bar{z})^{s+n}]}$
.

In studying the effect of \Eq{sp2} on two-particle states of well-defined total angular momentum $L$, one easily observes that $H$ annihilates all states  with $L<2s-1$. This is so because  all such  states  are proportional to at least a third power of $(z_1-z_2)$. (By rotational  invariance, it is sufficient to observe that all states with  total $L_z<2s-1$ have this property when either $z_1$ or $z_2$ are sent to the North pole at $z=0$.)
It further turns out that for two fermions in the lowest two Landau levels, there are  two representation with $L=2s+1$,  one representation with  $L=2s$, and three representations with $L=2s-1$, as one easily finds by focusing on highest weight states with $L=L_z$. The corresponding six highest weight states are, respectively,
\begin{equation}\label{sphere16}
\begin{matrix}
\ket 1=c_{0,s}^\dag c_{1,s+1}^\dag\ket 0,\\\\\ket 2=c_{0,s}^\dag c_{0,s-1}^\dag\ket 0,\\\\
\ket 3=(\sqrt{\frac{s}{1+2s}}c_{0,s}^\dag c_{1,s}^\dag-\sqrt{\frac{1+s}{1+2s}}c_{0,s-1}^\dag c_{1,s+1}^\dag)|0\rangle,\\\\
\ket 4=c_{1,s+1}^\dag c_{1,s}^\dag\ket 0,
\\
\\
\ket 5=(\sqrt{\frac{2s-1}{2(1+4s)}}c_{0,s}^\dag c_{1,s-1}^\dag-\sqrt{\frac{(4s^2-1)}{2s(1+4s)}}c^\dag_{0,s-1} c^\dag_{1,s}
\\
\\+\sqrt{\frac{(1+2s)(1+s)}{2s(1+4s)}}c_{0,s-2}^\dag c_{1,s+1}^\dag)\ket 0.
\\
\\
\ket 6=(\sqrt{\frac{1+s}{1+4s}}c_{1,s+1}^\dag c_{1,s-2}^\dag-\sqrt{\frac{3s}{1+4s}}c_{1,s}^\dag c_{1,s-1}^\dag)\ket 0,
\end{matrix}
\end{equation}
There is an obvious correspondence between the above six states and the  six states identified in Eqs. \eqref{Q16} for the disk geometry. Hence we expect that  there are still two zero modes contained in the subspace spanned by these six states, as happened in the disk geometry.
Taking into account the lower $L_z$ descendants of these states,
this will then lead to four non-zero energy two-particle  states for given $L_z=2R$, except for extremal values of $L_z$. Working first at the highest level,  one finds that there are two zero modes among the  $L=2s-1$ states $\ket{1}$, $\ket{5}$, and $\ket{6}$, and non-zero energy eigenstates
correspond  to the linear combinations

\begin{subequations}\label{sphereLC}

\begin{eqnarray}\nonumber&\ket{\tilde{1}}&=\frac{\sqrt{2}}{(17 s^2+6 s+1)^{1/4}\sqrt{s+1}}\\\nonumber
&\Big(&\frac{\sqrt{(s+1)\sqrt{17 s^2+6 s+1}-(s^2+4 s+1)}}{2}\ket 1+ \\&&
 \frac{s\sqrt{(2s+1)(2s+3)}}{\sqrt{(s+1)\sqrt{17 s^2+6 s+1}-(s^2+4 s+1)}}\ket 4\Big),
\end{eqnarray}
\begin{eqnarray}\nonumber
\ket{\tilde{2}}&=&-\frac{\sqrt{s(2 s+1)(4 s+1)}} {(s+1)\sqrt{6(6s-1)}}\ket 2 \\\nonumber
&+&\frac{ \sqrt{(2 s+1)(2 s-1)(2s+3)} }{(s+1)\sqrt{3(6s-1) }}\ket 5\\
&+&\frac{\sqrt{s}(2s+3)}{(s+1)\sqrt{2(6s-1)}}\ket 6 ,
\end{eqnarray}

\begin{eqnarray}\nonumber\\\nonumber&\ket{\tilde{4}}&=\frac{\sqrt{2}}{(17 s^2+6 s+1)^{1/4}\sqrt{s+1}}\\\nonumber
&\Big(&-\frac{\sqrt{(s+1)\sqrt{17 s^2+6 s+1}+(s^2+4 s+1)}}{2}\ket 1+\\&&
\frac{s\sqrt{(2s+1)(2s+3)}}{\sqrt{(s+1)\sqrt{17 s^2+6 s+1}+(s^2+4 s+1)}}\ket 4\Big),
\end{eqnarray}
\end{subequations} and $\ket{\tilde 3}=\ket{3}$,
with $L=2s-1$, $2s+1$, $2s+1$, and $2s$, respectively. This implies the following  form of the  $n=2$ Hamiltonian on the sphere,
\begin{equation}\label{HSphere}
\begin{split}
H&=\frac{1}{{4\pi }}\sum_{R\in\{-s-1,-s-\frac 12,\dotsc,s+1\}}\\ \Bigg(
&\frac{6 (2 s+1) (6 s-1)}{(16 s^2-1)}{T^{(2)\dagger}_R} T^{(2)}_R
+\frac{2(2s+3)}{4s+1}{T^{(3)\dagger}_R} T^{(3)}_R \\
\\
&+\frac{2(2 s+3) (-\sqrt{17 s^2+6 s+1}+5s+2)}{(4 s+1) (4 s+3)}{T^{(4)\dagger}_R} T^{(4)}_R \\
&+\frac{2(2 s+3) (\sqrt{17 s^2+6 s+1}+5s+2)}{(4 s+1) (4 s+3)}{T^{(1)\dagger}_R} T^{(1)}_R \Bigg),
\end{split}
\end{equation}
where we have also made explicit the eigenvalues corresponding to the eigenstates in \Eq{sphereLC},  and introduced  two-particle projection operators  
$T^{(i)\dagger}_RT^{(i)}_R$
onto  two-particle states
 $T^{(i)\dagger}_R\ket{0}$ that, at the appropriate highest weight value of $L_z$, correspond to the states $\ket{\tilde j}$, $j=1\dotsc 4$.
 To be more explicit, we first define
similar operators $Q^{(i)\dagger}_R$
that correspond in the same manner to the
two particle states $\ket{j}$, $j=1\dotsc 6$, \Eq{sphere16}:
\be\label{Qsphere}
\begin{split}
&Q^{(1)}_R=\sum_x
\braket{s,R+x;s+1,R-x|2s+1,2R} c_{1,R-x} c_{0,R+x}\\
&Q^{(2)}_R=\frac{1}{\sqrt{2}}\sum_x
\braket{s,R+x;s,R-x|2s-1,2R} c_{0,R-x} c_{0,R+x}\\
&Q^{(3)}_R=\sum_x
\braket{s,R+x;s+1,R-x|2s,2R} c_{1,R-x} c_{0,R+x}\\
&Q^{(4)}_R=\frac{1}{\sqrt{2}}\sum_x
\braket{s+1,R+x;s+1,R-x|2s+1,2R}\\ &\qquad\qquad\qquad c_{1,R-x} c_{1,R+x}\\
&Q^{(5)}_R=\sum_x
\braket{s,R+x;s+1,R-x|2s-1,2R}\\ &\qquad\qquad\qquad c_{1,R-x} c_{0,R+x}\\
&Q^{(6)}_R=\frac{1}{\sqrt{2}}\sum_x\braket{s+1,R+x;s+1,R-x|2s-1,2R}\\
& \qquad\qquad\qquad c_{1,R-x} c_{1,R+x}
\end{split}
\ee
Here, $\braket{j1,m1;j2,m2|j,m}$ is a Clebsch-Gordan coefficient. From \Eq{Qsphere}, we then form operators $T^{(i)}_R$ in a manner exactly as shown in \Eq{sphereLC}. We observe that the zero mode condition can still be cast in the form of \Eq{zero}.
It is further worth noting that in the limit  $s\rightarrow\infty$,  \Eq{HSphere} recovers the form of \Eq{H2} for the infinite disk geometry.

We are now in a perfect position to transcribe Lemmas 1-6 to the situation on the sphere. Upon reviewing the logic underlying the proofs of these lemmas, one finds that these hold generically for Hamiltonians of the form Eqs. \eqref{H2}, \eqref{HSphere}, provided that certain
coefficients at distances $|x|\leq 1$ are non-zero in the $Q$-operators, in this case \Eq{Qsphere}, as well as certain determinants involving these coefficients, which describe the linear relations used in the proofs of the lemmas. For the sphere, the relevant Clebsch-Gordan coefficients at $j_1-j_2-j \leq 3$ can be obtained from a standard sum\cite{racah1942theory,edmonds} that never has more than four terms, which especially for
small $|x|\leq 1$ are similar and can be combined into manageable closed forms. One thus verifies that the coefficients
of \Eq{Qsphere}
satisfy all the above mentioned non-vanishing conditions for Lemmas 1-6 to hold.
As a result, the only detail about these Lemmas that must be modified are the precise ratios in Lemma 5. Here we state this modified version:

{\textit{Lemma 5} (sphere)} If x0x appears in root patterns of a zero mode $\ket{\psi_0}$, then the proportions of coefficients of root patterns having !0!, !01, and 10! with all other occupancies the same are $2\sqrt{2s+3}$:$\sqrt{(s-j+2)(s+j)}$:$-\sqrt{(s+j+2)(s-j)}$, where $j$ is the angular momentum of the ``0'' in x0x.

Again we note that one recovers the proportions stated earlier for the disk geometry upon taking the limit $s,j\rightarrow \infty$ with $s-j$ finite.

Of course,  the new Lemma 5 does not change the zero mode counting on the sphere in terms of dominance patterns, for which the only relevant modification is the boundary condition discussed initially and in Fig. \ref{spherepattern}.  As explained, the above in particular confirms that the Jain-2/5 state satisfying
$2s=5/2N-{4}$ is the unique zero mode at this particular filling factor, with no zero modes existing at larger filling factor.
\section{Discussion\label{discussion}}

In the above we have established a description in terms of dominance patterns for the zero modes of the parent Hamiltonian of the unprojected Jain-2/5 state.
In doing so, we have further developed techniques to extract rules governing such patterns directly from a Hamiltonian principle. We found that, like in other examples\cite{SY1,SY2} where additional degrees of freedom beyond guiding centers are present, dominance patterns are not necessarily product states, but are subject to rules requiring simple entanglement under various circumstances. These rules may be thought of as further generalizations of conventional GPPs describing product states.
The rules we found
are nonetheless sufficiently simple to serve in zero mode counting, and we have in fact proven
that this procedure correctly gives the dimension of the zero mode space at given angular momentum and particle number. We have established this for both the disk and  spherical geometries, and demonstrated that zero mode counting at fixed angular momentum and particle number -- but with no restriction on quantum numbers describing relative occupancy of CF Landau levels or associated ``winding numbers'' in the effective edge theory -- is in agreement with the mode counting of the conformal field theory describing the edge physics.

The general approach followed in this paper
emphasizes the study of FQH parent Hamiltonians using second quantized methods in a context in which traditionally first quantized language has been given preference.
Indeed, only recently the second quantized presentation of FQH Hamiltonians has become a subject of interest in its own right.\cite{ortiz,Chen14,Mazaheri14, Thomale}
For one thing, it can be argued that this approach more readily gives access to spectral properties at finite energies.\cite{Amila15}
For another, the second quantized approach seems to be effective also in unravelling the zero mode structure of special Hamiltonians, as the present example demonstrates. We emphasize again that few examples seem to have been studied systematically in this regard where the wave function is not described by holomorphic polynomials, i.e., is not contained within the lowest Landau level. The advantage of our approach is that it directly ties the zero mode structure to a GPP for dominance patterns.
Such close ties between GPPs and Hamiltonians satisfying a zero mode paradigm may in fact explain why parent Hamiltonians have not been found in certain settings. For example, in the case of Jain states that are projected onto the lowest Landau level, the methods presented here strongly suggest that a parent Hamiltonian satisfying the zero mode paradigm would also lead to a GPP consistent with the effective edge theory. That is, to a set of rules governing the fusion of certain local building blocks on a one-dimensional lattice that leads to a densest possible state at the correct filling factor, and yields the correct zero mode counting at larger angular momenta. We conjecture that such a GPP is not possible for the Jain-2/5 state if the particles subject to the GPP have only the angular momentum (or guiding center) degrees of freedom of a single Landau level, with no additional degrees of freedom present(such as spin, Landau level indices, etc...). More generally, we conjecture that this is true for any state with an edge theory rich enough to comprise at least two branches of non-interacting chiral bosons: It appears that a ``plain vanilla'', single component GPP cannot be combinatorially rich enough to account for such edge theories.
On the other hand, how such GPPs are possible when additional degrees of freedom are present was seen here for the case of additional Landau level degrees of freedom. Similar, but distinct GPPs are implicit in Ref. \onlinecite{SY1} for, e.g., the (two-component) Halperin (332)-state, which has filling factor 2/5 but a different topological shift than the Jain-2/5 state. We leave the proof of this conjecture as a
challenge for future work.

It may be worth noting that, despite our emphasis on edge physics, there is no sharp distinction between edge and (quasi-hole type) bulk excitations from the point of view of dominance patterns.
This is of course expected in any microscopic theory, and is a consequence of the holographic principle. General bulk excitations in Abelian FQH states can be organized into a `lattice of excitations',\cite{read90} which is two-dimensional in the present case, and accommodates both charged and neutral excitations.
It is quite clear, e.g., that defects of the form $\dotsc$x0x00!00x0x$\dotsc$, $\dotsc$x0x00100x0x$\dotsc$, represent excitations of the same charge $1/5$, but differ by a neutral excitation. They would then have the same statistics.\cite{read90}
The results of the present paper also lay the basis to study such properties of bulk excitations, in particular pertaining to their statistics, in terms of dominance patterns using the coherent state method of earlier works.\cite{seidel_lee07,seidel_pfaffian,Flavin}

We point out that our results also  rigorously imply certain properties of the lowest LL {\em projected} Jain-2/5 state, and, more generally, CF states of the form \eqref{1qzm}.
On the sphere, e.g., all Slater determinants contributing to the projected Jain-2/5 state
must be obtainable via inward squeezing from the dominance pattern 100x0x00x0x$\dotsc$ x0x001. This pattern, of course, does by itself not appear in the projected Jain-2/5 state, as the first and last occupied orbital belong to the first excited LL. The projected Jain-2/5 state was studied from this point of view before in Ref. \onlinecite{RBH}, where a different dominance pattern was identified that becomes ``non-expandable'' in our terminology after projection.
The general pattern $\dotsc !0!00!0!00!0!\dotsc$ has also appeared in a thin torus study of the lowest LL projected Coulomb interaction.\cite{BK2}

While we have focused on the case of the Jain-2/5 parent Hamiltonian for definiteness,
the validity of our approach is certainly not limited to this case or those presented earlier along similar lines.\cite{ortiz,Chen14,Mazaheri14}
In particular, generalization to more than two-body Hamiltonians is certainly possible.
Even beyond the realm of FQH physics, attractive features of frustration free lattice Hamiltonians that are not necessarily finite ranged but feature a ``center-of-mass-conservation'' symmetry have long been advertised.\cite{LL,seidel05}
We are hopeful that the methods developed here will make major contributions to the general study of such Hamiltonians, the general $n$ case of \Eq{H} being a particular example.

\vspace*{+5mm}
\section{Conclusion\label{conclusion}}
In this paper, we have further developed a method to  extract GPPs governing zero modes of a FQH parent Hamiltonian directly from its second quantized form. In particular, we have demonstrated that such principles apply to states involving higher Landau levels, and provided an in-depth analysis of the zero mode structure of the Jain-2/5 state parent Hamiltonian and its realization through certain dominance patterns. As in earlier works focusing on single Landau level physics, we have identified single particle operators that generate zero modes. Our approach does, somewhat uncharacteristically, emphasize the second quantized presentation of parent Hamiltonians, which we developed in detail for the Jain-2/5 state for the disk and sphere geometries. The cylinder geometry can be treated similarly, with implications for the torus. This represents one route to a presentation of the physics that manifestly exposes the dynamics of the guiding centers and retains dynamical momenta only to the extent that they have not been eliminated by Landau level projection.
These aspects seem to be much in keeping with a line of thought recently put forth by Haldane.\cite{Haldane11}
A powerful strategy in exploring correlated electron physics is to stabilize special wave functions associated to certain fixed points
in the phase diagram via local Hamiltonians.
For the phases described by Jain states, lowest Landau level projected versions of Jain states, or manifestly projected hierarchy states, are sometimes thought to be the proper fixed point wave functions, since they are compatible with the strong field limit. We have presented arguments here why a local parent Hamiltonian for these states may not be possible, at least not if we want it to fall within the usual zero mode paradigm.
It is then reassuring that the existing parent Hamiltonian for the unprojected
Jain-2/5 state does fall into this paradigm, as we argued in great detail. The Hamiltonian studied here is the $n=2$ special case of a family of Trugman-Kivelson interactions projected onto $n$ Landau levels.
We expect that the methodology developed here will be of great value to shed light on the case of larger $n$. We leave this as an interesting problem for the future.

\begin{acknowledgments}
This work was supported in part by NSF grant
No. DMR-1206781. We are indebted to J. Jain for bringing Ref. \onlinecite{ReMac} to our attention. Further stimulating discussions with G. Ortiz, Z. Nussinov and K. Yang are gratefully acknowledged.
AS would moreover like to thank D.-H. Lee and J.M. Leinaas for insightful discussions on related issues.

\end{acknowledgments}
\bibliography{twofifth}\end{document}